\documentclass[showpacs, amsmath,amssymb, aps]{revtex4}

\usepackage{tabularx}
\usepackage{graphicx}
\usepackage{float}
\usepackage{dcolumn}
\usepackage{bm}
\usepackage{mathrsfs}
\usepackage{amsmath}

\begin{document}


\title{Light on curved backgrounds}

\author{D. Batic}
\email{davide.batic@uwimona.edu.jm}
\affiliation{%
Department of Mathematics,\\  University of the West Indies, Kingston 6, Jamaica 
}

\author{S. Nelson}
\email{stacyann.nelson@mymona.uwi.edu}
\affiliation{%
Departament of Physics,\\ University of the West Indies, Kingston 6, Jamaica
}
\author{M. Nowakowski}
\email{mnowakos@uniandes.edu.co}
\affiliation{
Departamento de Fisica,\\ Universidad de los Andes, Cra.1E
No.18A-10, Bogota, Colombia
}

\date{\today}

\begin{abstract}
We consider the motion of light 
on different spacetime manifolds by calculating
the deflection angle, lensing properties and by
probing into the possibility of bound states.
The metrics in which we examine
the light motion include, among other, a general
relativistic Dark Matter metric, 
a dirty Black Hole and a Worm Hole metric,
the last two inspired by non-commutative geometry.
The lensing in a Holographic Screen metric is 
discussed in detail.
We study also the bending of light
around naked singularities like, e.g.,
the Janis-Newman-Winicour metric
and include other cases.
A generic property of light
behaviour in these exotic metrics
is pointed out.  For the standard
metric like the Schwarzschild and Schwarzschild-de Sitter
cases we improve the accuracy of the lensing
results for the weak and strong regime.

\end{abstract}

\pacs{XXX}
\maketitle
\
\section{Introduction}
The year 2015 has been declared by UN and UNESCO
as the `` International Year of Light'' \cite{UN} which commemorates the
achievements of light sciences. 
Simultaneously, the year 2015 is the centenary year
of General Relativity. Light bending
is a genuine effect of General Relativity (at least
when we confront the theoretical prediction with
observations) and was
the first experimental confirmation of the newly discovered theory
which, looking back, is quite an achievement. 
It seems therefore timely to revive
the subject of light motion on
curved backgrounds by including
new interesting examples of recently emerged 
metrics and generalizing or improving/correcting
existing analytical results.

To appreciate the development of the subject let us recall that
the idea of light bending can be traced back to Newton's
\emph{Opticks} \cite{Newton} which concludes with a number of Queries.
In Query 1 \cite{Soares} Newton states:
\emph{``Do not Bodies act upon Light
at a distance, and by their action bend its Rays, and is not
this action (caeteris paribus) strongest at the least distance''}.
However, according to \cite{Will} the first 
concrete calculation within the Newtonian
framework was done by Henry Cavendish in 1784,
but the result remained unpublished. 
Twenty years later
Johann Georg von Soldner has done a similar calculation
\cite{Soldner} and taking into account that
both these calculations assume a different position of the 
light source (Cavendish light is emitted at infinity whereas Soldner's comes form a surface of a
gravitating body), both results agree in the first order
approximation \cite{Will}. Moreover, the result,
$\Delta \phi_N=2GM_{\bigodot}/R_{\bigodot}$ ($M_{\bigodot}$ and $R_{\bigodot}$ are the mass
and radius of the sun), is
half the value obtained from General Relativity using the
Schwarzschild metric. Einstein first attempt
to calculate the effect of gravity on light
yielded the same value obtained by Cavendish and Soldner
as he also used the Newtonian theory by invoking the
energy-mass equivalence. Only in 1916 he obtained
what is now considered the correct result (twice the Newtonian
value)
within the framework of General Relativity. Using the Robertson expansion of the
metric the deflection angle can be parametrized as
$\Delta \phi_N=4GM_{\bigodot}/r_0 (1+\gamma)/2$ ($r_0$ is the closest approach). The most precise
experiments on Quasars using very long baseline radio interference 
gives $(1+\gamma)/2=0.99992 \pm 0.00023$ \cite{Will2}, an impressive
result which excludes the Newtonian value $\gamma=1$ and confirms
General Relativity. Of course, the above mentioned
theoretical result in General Relativity is only the very
first approximation in the Schwarzschild metric.  
More accurate expansions are possible and will be presented in this paper.
The result is valid up to $(2M/r_0)^5$ based on expansion
of incomplete elliptic integrals of the first kind.

A more pronounced effect of light bending is gravitational lensing.
The idea is usually attributed to Einstein \cite{EinsteinL},
but has been already published twelve years earlier
by Chwolson \cite{Chwolson}. Since then the field
has achieved a remarkable level, from the observational as well
as from theoretical point of view \cite{reviews} 
reaching a high mathematical sophistication \cite{math}.
In spite of the seniority of the subject,
one can still find some niches where an improvement is possible.
As far as the lensing in the Schwarzschild metric is
concerned we offer some corrections of expressions
existing in the literature and generalization of the
formulae.

Another metric, closely related to the Schwarzschild metric,
is the Kottler or Schwarzschild-de Sitter metric which
includes a positive cosmological constant. 
The interest in this metric was revived after
the accelerated stage of the expansion of the Universe
was discovered and a positive cosmological constant
could account for the observational data. 
A natural question arises: does the
cosmological constant, given its ``measured value'' affect
the properties of light deflection? This led to some controversy
in the literature regarding the observability of the
cosmological constant in the lensing. 
Considering that with the inclusion of the cosmological
constant two different scales appear in the theory,
it is a priori not excluded that an effect 
combines these two scales in a way that it becomes,
in principle, measurable. This does not seem
to happen for the cosmological constant at least as far as lensing is concerned.
Our result contains
the cosmological constant, but the effect is tiny.
In deriving this result we made sure that our expressions
reduce to the formulae encountered in the Schwarzschild metric
when we put the cosmological constant to zero.

In the whole paper we probe into the light
motion around naked singularities. 
One case studied in detail is the
Janis-Newman-Winicour (JNW) metric.
One of the interest is to get an insight
of motion of light in naked singularities.
In the case of the JNW metric we concentrate on generalization
of existing results, but embed the JNW results
also in a wider context. With a simple qualitative tool
we can show that in many cases of naked singularities
(apart from  the JNW metric, we examine three other
naked singularities) a narrow range of possible
parameters leads to bound states of light.

The relevance of lensing is, of course, manifold.
But it is probably its connection to Dark Matter (DM) \cite{lensingDM}
which led the area to the avenues of 
new physics. Based on the non-relativistic Dark Matter
density of Navarro-Frankel-Simon a
general relativistic metric has been derived earlier which makes
it possible to study the deflection of light in the galactic
DM halos  from
the scratch. We offer here the first results of such a study.

Next we focus on a class of metrics which have been derived
having in mind some models of quantum gravity.
To be specific, we calculate the light deflection angle for two Black Hole (BH)
metrics: the holographic screen and the dirty BH inspired
by noncommutative geometry. The dirty BH contains in a limit the case
of a noncommutative BH when the parameters are chosen
accordingly. These metrics are based on the fact that noncommutativity,
i.e., $[x^{\mu}. x^{\nu}]=i\theta^{\mu \nu}$, leads to smeared objects
in place of point-like particles \cite{S}. 
On the other hand, the holographic screen metric
has been obtained by ``reversed engineering'' demanding that the metric
has no curvature singularity and ``self-implements'' 
the characteristic length
scale. 
It is, of course, of some interest to see how light
behaves in exotic metrics like the Worm Hole metric and
mini BH metrics which include aspects of noncommutative
geometry.

The choice of the metrics to study the light deflection
covers a wide range: from the standard general relativistic metrics
of BHs over a naked singularity and a DM galactic halo metric to metrics
inspired by some aspects of quantum gravity out of which one is a worm hole.
We hope to have added to the subject of light bending in
gravitational fields some new insight by studying these examples.

\section{Basic formulae}
We consider a static and radially symmetric gravitational field represented by the line element ($c=G=1$)
\begin{equation}\label{25.3}
ds^2=B(r)dt^2-A(r)dr^2-r^2 C(r)[d\vartheta^2+\sin^2{\vartheta}d\varphi^2] 
\end{equation}
with $r\in(0,+\infty)$, $\vartheta\in[-\pi,\pi]$, and $\varphi\in[0,2\pi)$. We further suppose that the functions $A$, $B$, and $C$ are at least 
$k$ times continuously differentiable on some interval $I\subseteq\mathbb{R}^{+}$. The corresponding relativistic Kepler problem is defined by the 
geodesic equation 
\[
\frac{d^2 x^\kappa}{d\lambda^2}= -\Gamma^\kappa_{\mu\nu}
\frac{dx^\mu}{d\lambda}\frac{dx^\nu}{d\lambda}
\]
for the metric (\ref{25.3}) and the relation
\begin{equation}\label{25.2}
g_{\mu\nu} \frac{dx^\mu}{d\lambda}\frac{dx^\nu}{d\lambda}=\left(\frac{ds}{d\lambda}\right)^2=\left(\frac{d\tau}{d\lambda}\right)^2=
\left\{ \begin{array}{ll}
         1 & \mbox{if}~m\neq 0;\\
         0   & \mbox{if}~m=0.
\end{array} \right.
\end{equation}
If we compute the Christoffel symbols we find that the geodesic equation gives rise to the following system of ordinary differential equations 
\begin{eqnarray}
\frac{d^2 x^0}{d\lambda^2}&=&-\frac{B^{'}(r)}{B(r)}\frac{dx^0}{d\lambda}\frac{dr}{d\lambda},\label{25.5}\\
\frac{d^2 r}{d\lambda^2}&=&-\frac{B^{'}(r)}{2A(r)}\left(\frac{dx^0}{d\lambda}\right)^2
-\frac{A^{'}(r)}{2A(r)}\left(\frac{dr}{d\lambda}\right)^2+\frac{r^2 C^{'}(r)+2rC(r)}{2A(r)}\left(\frac{d\vartheta}{d\lambda}\right)^2
+ \frac{r^2 C^{'}(r)+2rC(r)}{2A(r)}
\sin^2{\vartheta}\left(\frac{d\varphi}{d\lambda}\right)^2,\label{25.6}\\
\frac{d^2\vartheta}{d\lambda^2}&=&-\frac{rC^{'}(r)+2C(r)}{rC(r)}\frac{dr}{d\lambda}\frac{d\vartheta}{d\lambda}
+\sin{\vartheta}\cos{\vartheta}\left(\frac{d\varphi}{d\lambda}\right)^2,\label{25.7}\\
\frac{d^2\varphi}{d\lambda^2}&=&-\frac{rC^{'}(r)+2C(r)}{rC(r)}\frac{dr}{d\lambda}\frac{d\varphi}{d\lambda}-2\cot{\vartheta}
\frac{d\vartheta}{d\lambda}\frac{d\varphi}{d\lambda}.\label{25.8}
\end{eqnarray}
Equation (\ref{25.7}) can be solved by imposing that $\vartheta=\pi/2$. There is no loss in generality in introducing this condition since at a 
certain point in time the coordinate system can be rotated in such a way that $\vartheta=\pi/2$ and $d\vartheta/d\lambda=0$. Then, the coordinate and velocity vectors will belong to the 
equatorial plane $\vartheta=\pi/2$. This implies that $d^2\vartheta/d\lambda^2=0$ and hence $\vartheta(\lambda)=\pi/2$. As a consequence 
the whole trajectory will belong to the equatorial plane and equation (\ref{25.8}) becomes
\begin{equation}\label{25.10}
\frac{1}{r^2 C(r)}\frac{d}{d\lambda}\left(r^2 C(r)\frac{d\varphi}{d\lambda}\right)=0. 
\end{equation}
The above equation can be immediately integrated and we obtain
\begin{equation}\label{25.11}
 r^2 C(r)\frac{d\varphi}{d\lambda}=\ell=\mbox{const}.
\end{equation}
We can also motivate the choice $\vartheta=\pi/2$ and equation (\ref{25.11}) with the help of the isotropy of the problem at hand. This approach is 
usually adopted in the treatment of the nonrelativistic Kepler problem where the angular momentum $\ell$ is conserved because of the isotropy of the 
problem. Since the direction of $\ell$ is constant, we can choose the coordinate system in such a way that 
${\bf{e}}_z\parallel{\bf{\ell}}$. This is equivalent to require that $\vartheta=\pi/2$. Since the magnitude of $\ell$ is constant, 
equation (\ref{25.11}) will hold. Hence, the integration constant can be 
interpreted as the angular momentum per unit mass, i.e. $\ell=L/m$. Let us write (\ref{25.5}) in the form
\begin{equation}\label{25.12}
\frac{d}{d\lambda}\left[\ln{\frac{dx^0}{d\lambda}}+\ln{B(r)}\right]=0. 
\end{equation}
Integrating the above equation we obtain
\begin{equation}\label{25.13}
B(r)\frac{dx^0}{d\lambda}=F=\mbox{const}. 
\end{equation}
If  we set $\vartheta=\pi/2$ and use (\ref{25.11}) and (\ref{25.13}) in (\ref{25.6}), as a result we get the equation
\begin{equation}\label{25.14}
\frac{d^2 r}{d\lambda^2}+\frac{F^2 B^{'}(r)}{2A(r)B^2(r)}+\frac{A^{'}(r)}{2A(r)}
\left(\frac{dr}{d\lambda}\right)^2-\frac{\ell^2[rC^{'}(r)+2C(r)]}{2r^3 A(r)C^2(r)}=0. 
\end{equation}
Multiplication by $2A(r)(dr/d\lambda)$ yields
\begin{equation}\label{25.15}
\frac{d}{d\lambda}\left[A(r)\left(\frac{dr}{d\lambda}\right)^2+\frac{\ell^2}{r^2 C(r)}-\frac{F^2}{B(r)}\right]=0. 
\end{equation}
One more integration finally gives
\begin{equation}\label{25.16}
A(r)\left(\frac{dr}{d\lambda}\right)^2+\frac{\ell^2}{r^2 C(r)}-\frac{F^2}{B(r)}=-\epsilon=\mbox{const.} 
\end{equation}
This radial equation can be seen as the most important equation of motion since the angular motion is completely 
specified by (\ref{25.11}) and the condition $\vartheta=\pi/2$ whereas the connection between $t$ and $\lambda$ is fixed by (\ref{25.13}).  
If we integrate (\ref{25.16}) once more we obtain $r=r(\lambda)$ and if we substitute this function into (\ref{25.11}) and 
(\ref{25.13}), we get after integration $\varphi=\varphi(\lambda)$ and $t=t(\lambda)$. Elimination of the parameter $\lambda$ yields 
$r=r(t)$ and $\varphi=\varphi(t)$. Together with $\vartheta=\pi/2$ they represent the full solution of the problem. The involved integrals 
cannot be in general solved in terms of elementary functions. To determine $\epsilon$, we rewrite (\ref{25.2}) as 
\begin{equation}\label{25.17}
g_{\mu\nu}\frac{dx^\mu}{d\lambda}\frac{dx^\nu}{d\lambda}=
B(r)\left(\frac{dx^0}{d\lambda}\right)^2- 
A(r)\left(\frac{dr}{d\lambda}\right)^2-
r^2 C(r)\left(\frac{d\vartheta}{d\lambda}\right)^2-
r^2\sin^2{\vartheta}~ C(r)\left(\frac{d\varphi}{d\lambda}\right)^2=
\epsilon,
\end{equation}
where in the last inequality we used the condition $\vartheta=\pi/2$ together with (\ref{25.11}), (\ref{25.13}) and 
(\ref{25.16}). From (\ref{25.17}) and (\ref{25.2}) it follows that
\begin{equation}\label{25.18}
\epsilon= \left\{ \begin{array}{ll}
         1 & \mbox{if}~ m\neq 0;\\
         0  & \mbox{if}~ m=0.\end{array} \right. 
\end{equation}
We want to determine the trajectory $\varphi=\varphi(r)$ in the equatorial plane 
$\vartheta=\pi/2$. First of all, we observe that (\ref{25.16}) gives
\begin{equation}\label{25.19}
\left(\frac{dr}{d\lambda}\right)^2=\frac{1}{A(r)}\left[\frac{F^2}{B(r)}-\frac{\ell^2}{r^2 C(r)}-\epsilon\right]. 
\end{equation}
Taking into account that $d\varphi/d\lambda=(d\varphi/dr)(dr/d\lambda)$ and using (\ref{25.11}) together with (\ref{25.19}), we obtain
\begin{equation}\label{25.20}
\left(\frac{d\varphi}{dr}\right)^2=\frac{A(r)B(r)}{r^4 C(r)}\left[\frac{F^2}{\ell^2}C(r)-\frac{B(r)}{r^2}-\frac{\epsilon}{\ell^2}B(r)C(r)\right]^{-1} 
\end{equation}
and integration yields
\begin{equation}\label{25.21}
\varphi(r)=\pm\int\frac{dr}{r^2}\sqrt{\frac{A(r)B(r)}{C(r)}} 
\left[\frac{F^2}{\ell^2}C(r)-\frac{B(r)}{r^2}-\frac{\epsilon}{\ell^2}B(r)C(r)\right]^{-1/2}+\widetilde{C}
\end{equation}
where $\widetilde{C}$ is an arbitrary integration constant. The plus and minus sign must be chosen for particles approaching the gravitational object on the 
equatorial plane and having trajectories exhibiting an anticlockwise and clockwise direction, respectively. This integral determines the trajectory 
$\varphi=\varphi(r)$ in the plane where the motion takes place. In the case of a massive particle ($\epsilon=1$) the trajectory depends on two 
integration constants ($F$ and $\ell$). In the case of a scattering problem these constants can be expressed in terms of an impact parameter and 
the initial velocity $r^{'}(r_0)$ where $r_0$ is the distance of closest approach to the gravitational object attained for some value $\lambda_0$ 
of the affine parameter. For massless particles ($\epsilon=0$) the trajectory depends only on the integration constant $F/\ell$ that can be 
interpreted as an impact parameter as follows \cite{Weinberg}
\[
\frac{1}{b^2}=\frac{F^2}{\ell^2}=\frac{B(r_0)}{r_0^2 C(r_0)}. 
\]
This in turn permits to express (\ref{25.21}) as 
\[
\varphi(r)=\pm\int\frac{dr}{r}\sqrt{\frac{A(r)}{C(r)}} 
\left[\left(\frac{r}{r_0}\right)^2\frac{C(r)B(r_0)}{C(r_0)B(r)}-1\right]^{-1/2}+\widetilde{C}. 
\]
It is useful to derive an inequality which gives us information for the qualitative orbits of the particle. It is straightforward to see that from 
(\ref{25.21}) we obtain the condition
\begin{equation} \label{effpot}
F^2>\frac{\ell^2 B(r)}{r^2 C(r)}+\epsilon B(r)\equiv V(r). 
\end{equation}
For massless particles we can write
\begin{equation}\label{vtilde}
\frac{F^2}{\ell^2}> \frac{B(r)}{r^2 C(r)}\equiv\widetilde{V}(r).
\end{equation}
As a first application of equation (\ref{vtilde}) consider the
Reissner-Nordstr\"om metric with
\[
B(r)=1-\frac{1}{r_s}+ \frac{r_Q^2}{r^2}
\]
where $r_s$ is the standard Schwarzschild radius and $r_Q$ proportional the
electric charge. We have a naked singularity if $r_s < 2r_Q$.
It is easy to show that $\widetilde{V}$ has a local minimum
and maximum if $\alpha\equiv r_s^2/r_Q^2 > 32/9$. Such a minimum in which
the photons would be trapped has a physical singificance only
if it occurs at a value bigger than the horizon $r_+$
(the horizons are $r_{\pm}=1/2/(r_s \pm\sqrt{r_s^2-4r_Q^2 })$
or in the case of a naked singularity.
In figure 1 we demonstrate that this minimum is of relevance only
if we have a naked singularity. Hence the range for such 
meaningful minimum
is very narrow, namely
\[
4> \alpha  > \frac{32}{9}
\]
If a particle in classical mechanics has a bound orbit (like in the
minimum) we usually talk about an attractive potential.
By the same token using similar nomenclature
we could claim that the existence of
photon's bound orbits around a naked singularity
is an indication for its attractive nature for light in
contrast to some other results for massive particles \cite{us1}.
We will see later that this statement is generic
for other naked singularities as well. 

\begin{figure}\label{RN}
\includegraphics[scale=0.5]{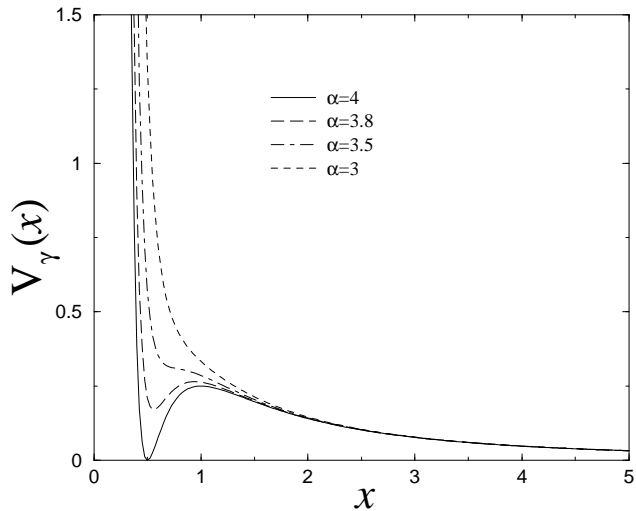}
\caption{\label{fig3}
Behaviour of the ``photon potential'' proportional to $\widetilde{V}$ for the Reissner-Nordstr\"om
naked singularity. As explained in the text in the narrow range of $\alpha$ there is
a possibility to trap photons inside a well.
}
\end{figure}

\subsection{Lens equation for compact gravitational objects}
We give a short derivation of the lens equation in the presence of compact gravitational objects, i.e. any astrophysical object whose size is comparable 
to the event horizon of a black hole. The corresponding lens equation may be also applied to black holes and other compact objects that 
did not fully undergo the gravitational collapse. Let S, L, O, and I denote the light source, the lens, the observer, and the image of the source seen 
by the observer, respectively. By OL we denote the optical axis. Furthermore, we introduce angles $\beta$ and $\theta$ giving the position of S 
with respect to OL and the position of I as seen by O, respectively. In general, the closest approach distance $r_0$ does not need to be identified 
with the impact parameter $b$. By $\Delta\varphi$ we denote the deflection angle. Then, simple trigonometric arguments lead to the full lens equation 
\cite{Weinberg,Bozza1,ellis,Schneider}
\begin{equation}\label{Il}
\tan{\beta}=\tan{\theta}-\frac{D_{LS}}{D_{OS}}\left[\tan{\theta}+\tan{(\Delta\varphi-\theta)}\right], 
\end{equation}
where $D_{LS}$ is the distance between $L$ and $S$, and $D_{OS}=D_{OL}+D_{LS}$ with $D_{OL}$ the distance between $O$ and $L$. Given $\beta$ and 
$\Delta\varphi$, (\ref{Il}) allows to compute the positions $\theta$ of the images I of S seen by O. The magnification of an image for circularly 
symmetric gravitational lenses is
\begin{equation}\label{mag}
\mu=\left(\frac{\sin{\beta}}{\sin{\theta}}\frac{d\beta}{d\theta}\right)^{-1}, 
\end{equation}
where the sign of $\mu$ controls the parity of the image. Critical curves are singularities of $\mu$ in the lens plane and the corresponding values 
in the source plane are called caustics. By images of 0-parity we mean critical images. The tangential and radial magnifications are given by
\begin{equation}\label{tr}
\mu_t= \left(\frac{\sin{\beta}}{\sin{\theta}}\right)^{-1},\quad
\mu_r= \left(\frac{d\beta}{d\theta}\right)^{-1}.
\end{equation}
Tangential and radial critical curves are simply singularities of $\mu_t$ and $\mu_r$, respectively. Their values on the source plane are called 
tangential and radial caustics. If $\beta$, $\theta$, and $\Delta\varphi$ are small (weak gravitational field), the tangent functions appearing in 
(\ref{Il}) can be Taylor expanded and the deflection angle becomes $\Delta\varphi=4M/r_0$ so that (\ref{Il}) can be solved producing two images 
whose separation from OL is quantified by the so-called Einstein angle
\begin{equation}\label{Einstein}
\theta_E=\sqrt{\frac{4M D_{LS}}{D_{OS}D_{OL}}}. 
\end{equation}

\section{Bending of light in the Schwarzschild metric}
Light rays experience a bending effect due to the presence of a gravitational field. We quantify this 
effect in the case of a lens represented by a Schwarzschild black hole with
\[
B(r)=1-\frac{2M}{r}, 
\]
where $M$ denotes the mass of the black hole. We will also suppose that the source, lens, and observer lie along a straight line. Since the 
Schwarzschild spacetime is asymptotically flat we will assume that the source and observer are located in the flat spacetime region. Setting 
$\epsilon=0$ in (\ref{25.21}) and taking into account that $AB=1$, we find that 
\begin{equation}\label{26.1}
\varphi(r)=\varphi(r_0)+\int_{r_0}^\infty\frac{dr}{r^2}
\left[\frac{F^2}{\ell^2}-\frac{B(r)}{r^2}\right]^{-1/2},
\end{equation}
where the source has been placed in the asymptotically flat region and as a starting point for the integration the minimal distance $r_0$ of the 
light ray from the surface of the gravitational object has been chosen. Note that the choice of a plus sign in front of the above integral 
corresponds to the fact that we are considering light rays moving on trajectories having an anticlockwise direction. Without loss of generality 
we also require that $\varphi(r_0)=0$. From $r=r_0$ to $r=\infty$ the angle $\varphi$ changes by a quantity $\varphi(\infty)$. Along the photon 
trajectory the radial vector undergoes a rotation with angle $2\varphi(\infty)$. If the gravitational object would be absent, we would have a 
straight line for the photon trajectory implying that $2\varphi(\infty)=\pi$. Hence, the angle by which light is bended by a spherically symmetric 
gravitational field described by an asymptotically flat metric is given by \cite{Weinberg,Fliessbach}
\begin{equation}\label{26.2}
\Delta\varphi=2\varphi(\infty)-\pi,\quad\varphi(r_0)=0. 
\end{equation}
Since in general $A(r)\neq 1$ the underlying three dimensional space is not Euclidean. For very large distances we have 
$A\to 1$ and $B\to 1$. This means that asymptotically far away from the gravitational object the light ray can be 
described as a straight line in the Euclidean space. The distance of closest approach is determined by the condition $r^{'}(r_0)=0$ with 
$r_0=r(\lambda_0)$ and we find that
\begin{equation}\label{26.4}
\frac{1}{b^2}=\frac{F^2}{\ell^2}=\frac{B(r_0)}{r_0^2},\quad b^2=\frac{r_0^3}{r_0-2M}. 
\end{equation}
This allows us to eliminate the constant $F^2/\ell^2$ in (\ref{26.1}) so that we can rewrite (\ref{26.2}) as 
\begin{equation}\label{26.5}
\frac{\Delta\varphi+\pi}{2}=\int_{r_0}^\infty\frac{dr}{r^2}
\left[\frac{B(r_0)}{r_0^2}-\frac{B(r)}{r^2}\right]^{-1/2}. 
\end{equation}
Depending on the values of the impact parameter we have the following scenarios \cite{Weinberg,Fliessbach,Darwin}
\begin{enumerate}
 \item 
if $b<3\sqrt{3}M$, the photon is doomed to be absorbed by the black hole;
\item
if $b>3\sqrt{3}M$, the photon will be deflected and it can reach spatial infinity. Here, we must consider two further cases
\begin{enumerate}
\item
if $b\gg 3\sqrt{3}M$, the orbit is almost a straight line and the deflection angle is approximately given by $4M/r_0$. Weak gravitational lensing 
deals with this case which corresponds to the situation when the distance of closest approach is much larger than the radius $r_\gamma$ of the 
photon sphere. For spherically symmetric and static spacetimes $r_\gamma$ can be computed by solving the equation \cite{Bozza2}
\begin{equation}\label{photon_sphere}
\frac{B^{'}(r)}{B(r)}=\frac{2}{r}. 
\end{equation}
In the case of the Schwarzschild metric we obtain $r_\gamma=3M$.
\item
If $0<b\ll 3\sqrt{3}M$, we are in the regime of strong gravitational lensing corresponding to a distance of closest approach $r_0\approx r_\gamma$. 
In this case the photon can orbit several times around the black hole before it flies off. 
\end{enumerate}
\end{enumerate}
Let $r_S=2M$ denote the Schwarzschild radius. As in \cite{Bozza1,Bozza2,SH} we rescale the time and radial coordinates as $\sigma=t/r_S$ and 
$\rho=r/r_S$. Then, the Schwarzschild radius is at $\rho_S=1$, the distance of closest approach will be given by $\rho_0=r_0/r_S$, and the radius 
of the photon sphere is located at $\rho_1=3/2$. Clearly, we must require that $\rho_0>\rho_1$. In terms of $\rho$ (\ref{26.5}) becomes 
\begin{equation}\label{phi1}
\Delta\varphi(\rho_0)=-\pi+2\int_{\rho_0}^\infty\frac{d\rho}{\rho^2}\left[\frac{B(\rho_0)}{\rho_0^2}-\frac{B(\rho)}{\rho^2}\right]^{-1/2}. 
\end{equation}
Let $y=B(\rho)$ and $z=(y-y_0)/(1-y_0)$ where $y_0=B(\rho_0)$. We find that (\ref{phi1}) can be written as
\begin{equation}\label{bozza_int}
\Delta\varphi(\rho_0)=-\pi+2\int_0^1 f(z,\rho_0)~dz,\quad
f(z,\rho_0)=\left[\left(2-\frac{3}{\rho_0}\right)z+\left(\frac{3}{\rho_0}-1\right)z^2-\frac{z^3}{\rho_0}\right]^{-1/2}
\end{equation}
The function $f$ has three singularities located at
\[
z_1=0,\quad z_2=\frac{3-\rho_0 -\sqrt{\rho_0^2+2\rho_0-3}}{2} ,\quad
z_3=\frac{3-\rho_0 +\sqrt{\rho_0^2+2\rho_0-3}}{2}.
\]
It is straightforward to verify that $z_2<0$, $z_3>\rho_1>0$, and $z_2<z_1<z_3$. If we factorize the argument of the square root in the expression 
for $f$ in terms of its roots and introduce the variable transformation $z=\widetilde{y}^2+z_2$ \cite{meech}, we obtain
\[
\Delta\varphi(\rho_0)=-\pi+4\sqrt{\rho_0}
\int_{\sqrt{-z_2}}^{\sqrt{1-z_2}}\frac{d\widetilde{y}}{\sqrt{(\widetilde{y}^2+z_2)(z_3-z_2-\widetilde{y}^2)}}. 
\]
Let $\widetilde{y}^2=z_2(1-k^2\sin^2{\phi})$ with $k=\sqrt{z_3/(z_3-z_2)}$. Then, 
\begin{equation}\label{esatto1}
\Delta\varphi(\rho_0)=
-\pi+A(\rho_0)F(\phi_1,k),\quad
A(\rho_0)=\frac{4\sqrt{\rho_0}}{z_3-z_2},\quad
\phi_1=\sin^{-1}{\sqrt{\frac{z_3-z_2}{z_3(1-z_2)}}},
\end{equation}
where $F$ denotes the incomplete elliptic integral of the first kind. Using the expansion $902.00$ in \cite{Byrd} for the incomplete elliptic integral of the first kind when $k\ll 1$, we find that the angle by which light is bended in a weak gravitational field ($\rho_0\gg 1$) is given by
\begin{equation}\label{wfl}
\Delta\varphi(\rho_0)=2\rho_0^{-1}+\left(\frac{15}{16}\pi-1\right)\rho_0^{-2}+\left(\frac{61}{12}-\frac{15}{16}\pi\right)\rho_0^{-3}
+\left(\frac{3465}{1024}\pi-\frac{65}{8}\right)\rho_0^{-4} 
+\left(\frac{7783}{320}-\frac{3465}{512}\pi\right)\rho_0^{-5}+\mathcal{O}(\rho_0^{-6}). 
\end{equation}
Taking into account that $\rho_0=r_0/(2M)$ the weak field approximation (\ref{wfl}) 
reproduces correctly the first order term $4M/r_0$ derived in \cite{Weinberg,Fliessbach,Darwin} and generalizes the weak field approximation 
derived in \cite{ellis,virba}. Moreover, it agrees with equation $(23)$ in \cite{keeton}. Concerning the strong deflection limit ($\rho_0\to\rho_1$) we will first show that the method used 
by \cite{Bozza2} is mathematically flawed and then we will use an asymptotic formula for the incomplete elliptic integral of the first kind 
derived in \cite{carlson}. First of all, \cite{Bozza2} starts by observing that the integrand $f$ appearing in the integral giving the deflection angle diverges as $z\to 0$. The order of divergence of the integrand can be found by expanding the argument of the square root in $f$ to the second order at $z=0$, more precisely
\[
f(z,\rho_0)\approx f_0(z,\rho)=\frac{1}{\sqrt{\alpha z+\beta z^2}},\quad\alpha= 2-\frac{3}{\rho_0},\quad \beta=\frac{3}{\rho_0}-1.
\]
Hence, for $\alpha\neq 0$ the leading order of the divergence of $f_0$ is $z^{-1/2}$ which can be integrated while for $\alpha=0$ the function $f_0$ diverges as $z^{-1}$ thus leading to a logarithmic divergence. 
\cite{Bozza2} splits the integral in (\ref{bozza_int}) as follows 
\[
2\int_0^1 f(z,\rho_0)~dz=I_D(\rho_0)+I_R(\rho_0),
\]
where
\[
I_D(\rho_0)=2 \int_0^1 f_0(z,\rho_0)~dz
\]
contains the divergence and 
\[
I_R(\rho_0)=2 \int_0^1 g_0(z,\rho_0)~dz,\quad g_0(z,\rho_0)=f(z,\rho_0)-f_0(z,\rho_0)
\]
is the original integral with the divergence subtracted. At this point one solves the above integrals and the sum of their results will give the deflection angle. The integral $I_D$ can be solved exactly and we get
\[
I_D(\rho_0)=\frac{4}{\sqrt{\beta}}\ln{\frac{\sqrt{\alpha+\beta}+\sqrt{\beta}}{\sqrt{\beta}}}
=-2\ln{\left(\frac{\rho_0}{\rho_1}-1\right)} +2\ln{2}+\mathcal{O}\left(\rho_0-\rho_1\right)     
\]
after having expanded $\alpha$ and $\beta$ around the radius of the photon sphere. To compute the residual integral $I_R$ \cite{Bozza2} employs the following expansion
\begin{equation}\label{regular}
I_R(\rho_0)=\sum_{n=0}^{\infty}\frac{(\rho_0-\rho_1)^n}{n!}\int_0^1\left.\frac{\partial^n g}{\partial\rho_0^n}\right|_{\rho_0=\rho_1}~dz
\end{equation}
and at the first order we find
\[
I_R(\rho_0)=\int_0^1 g(z,\rho_1)~dz+ \mathcal{O}\left(\rho_0-\rho_1\right)=2\ln{6(2-\sqrt{3})}+\mathcal{O}\left(\rho_0-\rho_1\right).
\]
\cite{Bozza2} claims that (\ref{regular}) can be used to compute all coefficients in the expansion for the regular part of the integral $I_R$. 
This is not true since a closer inspection of the partial derivatives $\partial^n g(z,\rho_1)/\partial\rho_0^n$ shows that they have the following  
behaviours as $z\to 0$ 
\[
\left.\frac{\partial g}{\partial\rho_0}\right|_{\rho_0=\frac{3}{2}}=-\frac{4}{3z}-\frac{2}{9}+\mathcal{O}(z),\quad
\left.\frac{\partial^2 g}{\partial\rho_0^2}\right|_{\rho_0=\frac{3}{2}}=\frac{40}{9z^2}-\frac{4}{27z}+\mathcal{O}(1),\quad
\left.\frac{\partial^3 g}{\partial\rho_0^3}\right|_{\rho_0=\frac{3}{2}}=-\frac{560}{27z^3}+\mathcal{O}(z^{-2}).
\]
Hence, the regular part of the integral giving the deflection angle cannot be represented by means of (\ref{regular}) because otherwise each 
coefficient with $n\geq 1$ in the expansion for $I_R$ would blow up due to the fact that $\partial^n g(z,\rho_1)/\partial\rho_0^n$ is never 
integrable at $z=0$ for all $n\geq 1$. We can overcome this problem by observing that $\sin{\phi_1}$ and $k$ both approach one 
as $\rho_0\to\rho_1$ since
\begin{equation}\label{kexp}
\sin{\phi_1}(\rho_0)=1-\frac{2}{9}(\rho_0-\rho_1)+\mathcal{O}(\rho_0-\rho_1),\quad
k(\rho_0)=1-\frac{4}{9}(\rho_0-\rho_1)+\mathcal{O}(\rho_0-\rho_1).
\end{equation}
For $\sin{\phi_1}$ and $k$ approaching one simultaneously the following asymptotic formula for the 
incomplete elliptic integral of the first kind holds \cite{carlson}
\begin{equation}\label{expansion}
F(\phi_1,k)=\frac{\sin\phi_1}{4}
\left\{[6-(1+k^2)\sin^2{\phi_1}]
\ln{\frac{4}{\cos{\phi_1}+\Delta}}-2+(1+k^2)\sin^2{\phi_1}+\Delta\cos{\phi_1}
\right\}+\theta F(\phi_1,k)
\end{equation}
with $\Delta=\sqrt{1-k^2\sin^2{\phi_1}}$ and relative error bound
\[
\frac{9\Delta^4\ln{\Delta}}{64\ln{(\Delta/16)}}<\theta<\frac{3}{8}\Delta^4.
\]
Taking into account that
\[
A(\rho_0)=4-\frac{8}{9}(\rho_0-\rho_1)+\mathcal{O}(\rho_0-\rho_1)
\]
and expanding (\ref{expansion}) around $\rho_0=\rho_1$ we finally obtain
\begin{equation}\label{sfl}
\Delta\varphi(\rho_0)=-\pi+\ln{144(7-4\sqrt{3})}-2\ln{\left(\frac{\rho_0}{\rho_1}-1\right)}+
\frac{16}{9}(\rho_0-\rho_1)+\mathcal{O}(\rho_0-\rho_1)^2.
\end{equation}
In order to express the deflection angle as a function of $\theta$ we must first rewrite $\rho_0$ in terms of the impact parameter. Taking into account that
\begin{equation}\label{ip_sch}
\widetilde{b}=\sqrt{\frac{\rho_0^3}{\rho_0-1}}=\widetilde{b}_{cr}+\sqrt{3}\left(\rho_0-\rho_1\right)^2
+\mathcal{O}\left(\rho_0-\rho_1\right)^3,\quad\widetilde{b}_{cr}=\frac{3\sqrt{3}}{2}
\end{equation}
we obtain the approximated relation
\begin{equation}\label{punto}
\rho_0-\rho_1\approx\sqrt{\frac{\widetilde{b}-\widetilde{b}_{cr}}{\sqrt{3}}}. 
\end{equation}
Substituting $\widetilde{b}=\theta\widetilde{D}_{OL}$ with $\widetilde{D}_{OL}=D_{OL}/r_S$ in (\ref{punto}) and replacing this relation in 
(\ref{sfl}) we finally get  
\begin{equation}\label{**}
\Delta\varphi(\theta)=-\pi+\ln{216(7-4\sqrt{3})}-\ln{\left(\frac{\theta\widetilde{D}_{OL}}{\widetilde{b}_{cr}}-1\right)}+ 
\frac{16}{9\sqrt{3}}\sqrt{\theta\widetilde{D}_{OL}-\widetilde{b}_{cr}}+\mathcal{O}(\theta\widetilde{D}_{OL}-\widetilde{b}_{cr})
\end{equation}
Our formula (\ref{sfl}) generalizes the Schwarzschild deflection angle in the strong field limit given by \cite{Bozza1,Darwin,Bozza2}. At this point a couple of remarks are in order. An expression for the exact deflection 
angle of photons in terms of elliptic integrals was first given in \cite{Darwin} (see equation $(29)$ therein), an equivalent representation is offered 
by \cite{Bozza1}, whereas \cite{Kogan} derives the deflection angle by applying formula $3.131.(5)$ at page $254$ in \cite{Gradsh} to our (\ref{bozza_int}). 
Furthermore, equation $(9)$ in \cite{Bozza1} expressing the modulus of the elliptic function is not correct and should read in the notation therein
\[
\lambda=\sqrt{\frac{3-x_0-\sqrt{-3+2x_0+x_0^2}}{3-x_0+\sqrt{-3+2x_0+x_0^2}}}.
\]
In Figure~2 we compare the analytical expression (\ref{esatto1}) of the deflection angle with (\ref{wfl}) and formula $(24)$ 
with $q=0$ in \cite{virba}.
\begin{figure}\label{F3}
\includegraphics[scale=0.5]{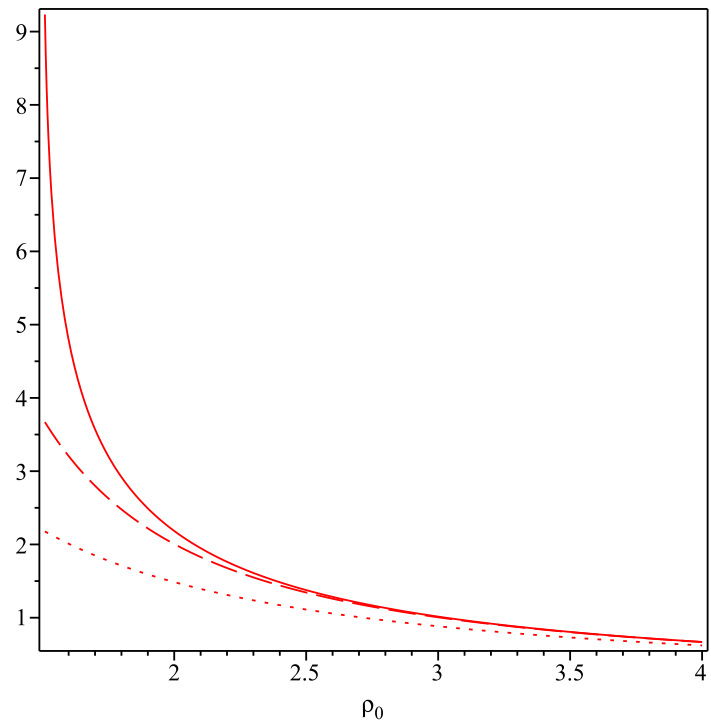}
\caption{\label{fig3.1}
Behaviour of the exact solution (\ref{esatto1}) (solid) versus the weak field approximations (\ref{wfl}) (dashed) and $(24)$ in \cite{virba} with $q=0$ (dotted).
}
\end{figure}
Figure~\ref{fa2} displays the exact solution (\ref{esatto1}) with the approximated solution 
$(50)$ offered by \cite{Bozza2} and our (\ref{sfl}). Figure~\ref{error} shows that using our expansion up to $\rho_0=1.55$ the error we introduce is about $0.1\%$ whereas the 
error committed by \cite{Bozza1} is $1\%$. 
\begin{figure}\label{Fa2}
\includegraphics[scale=0.5]{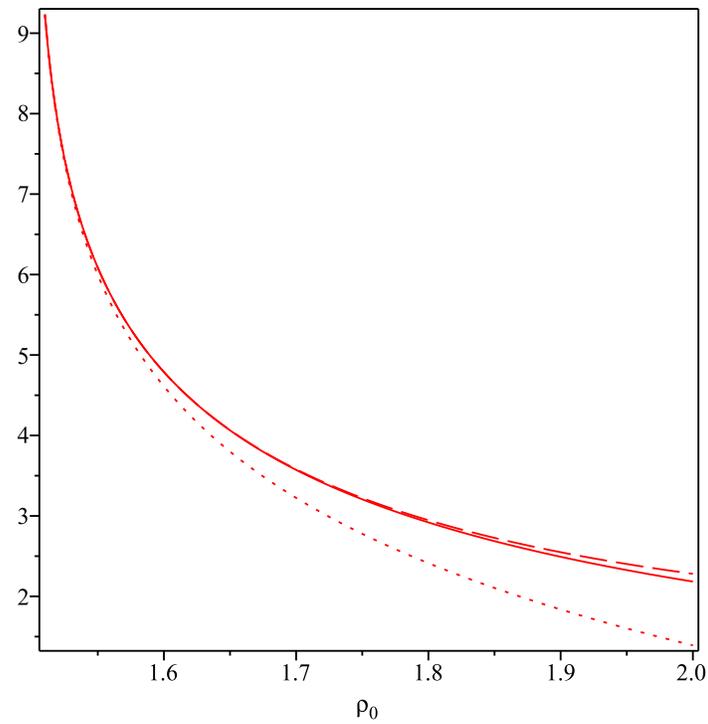}
\caption{\label{fa2}
Plot of the exact solution (\ref{esatto1}) (solid), Bozza's approximation (dotted) and the approximation (\ref{sfl}) (dashed). 
}
\end{figure} 
\begin{figure}\label{errorB}
\includegraphics[scale=0.5]{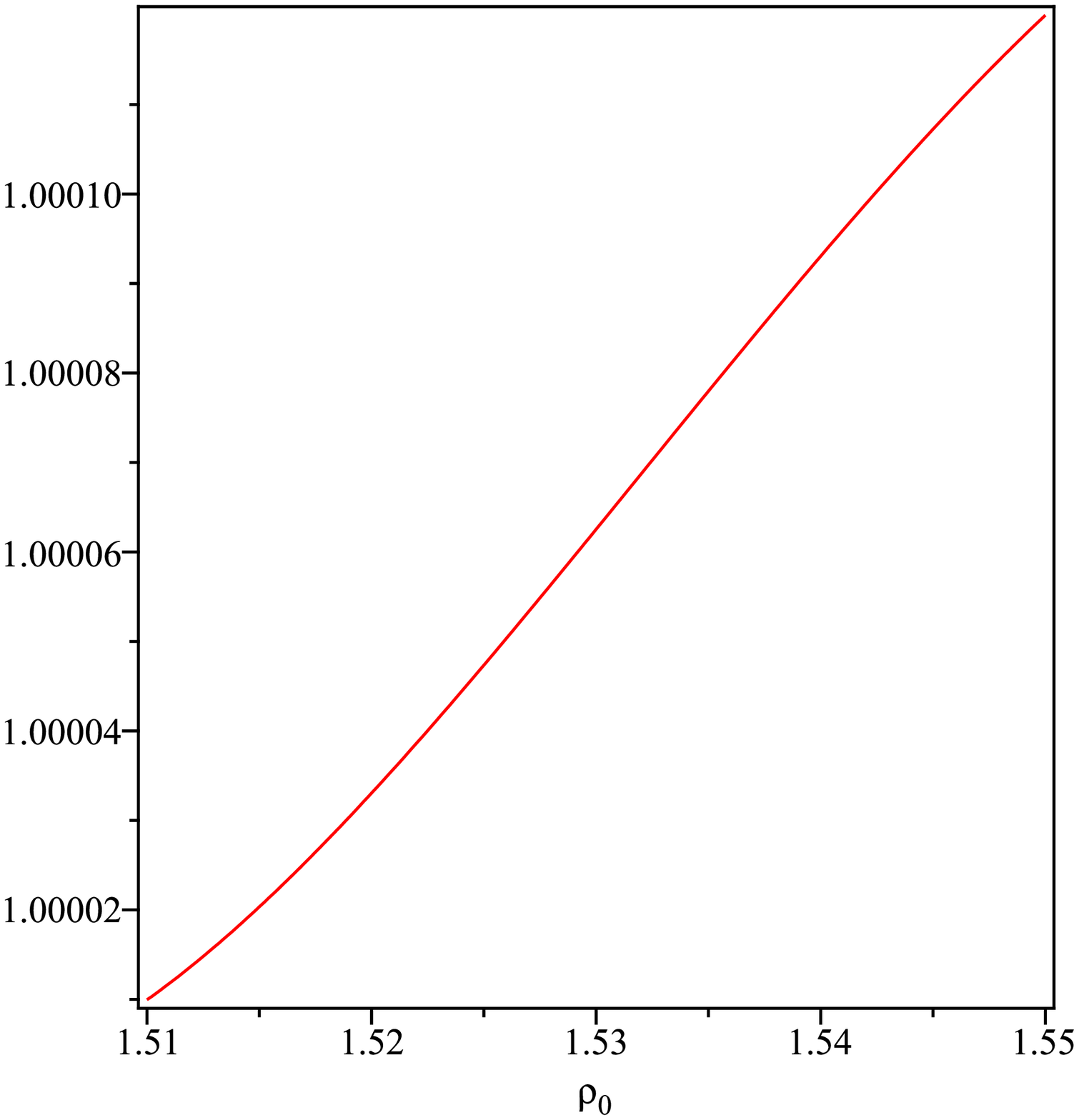}
\caption{\label{error}
Ratio of the exact deflection angle (\ref{esatto1}) and the approximate one (\ref{sfl}) as a function of the closest approach distance. 
}
\end{figure}
In the strong field approximation, when the light ray gets closer and closer to the photon sphere, $\Delta\varphi$ may become bigger than $2\pi$. This implies that the light ray will wind around the lens one or more times before escaping the gravitational pull. In the extreme case 
$\widetilde{b}_{cr}=3\sqrt{3}/2$ corresponding to $\rho_0=3/2$, $\Delta\varphi$ diverges logarithmically and the photon is captured by the photon sphere. The approximated lens equation in this regime is 
given by \cite{Bozza1}
\begin{equation}\label{II}
\beta=\theta-\frac{\widetilde{D}_{LS}}{\widetilde{D}_{OS}}\Delta\varphi_n,\quad\Delta\varphi_n=\Delta\varphi-2n\pi,\quad
n\in\mathbb{N}. 
\end{equation}
Let $\theta^0_n$ denote the values of $\theta$ such that 
$\Delta\varphi_n(\theta^0_n)=0$, or equivalently $\Delta\varphi(\theta^0_n)=2n\pi$. Using (\ref{**}) this equation can be solved for $\theta^0_n$ and we obtain
\begin{equation}\label{box}
\theta^0_n=\frac{1}{\widetilde{D}_{OL}}
\left[\frac{3\sqrt{3}}{2}+\frac{243}{64}W_n^2\right],\quad W_n=W\left(-16\sqrt{7-4\sqrt{3}}e^{-\pi\left(n+\frac{1}{2}\right)}\right)
\end{equation}
where $W_n$ denotes the Lambert function \cite{corless}. Our (\ref{box}) generalizes formula $(16)$ in \cite{Bozza1}. In the limit $n\to\infty$, i.e. when the photon makes an infinite number of loops around the black hole we correctly obtain 
$\theta_\infty^0=(3\sqrt{3})/(2\widetilde{D}_{OL})$. If we expand $\Delta\varphi$ at the first order around $\theta_n^0$, (\ref{**}) gives
\begin{equation}\label{***}
\Delta\varphi_n=\Delta\varphi-\Delta\varphi(\theta^0_n)=\left.\frac{d\Delta\varphi}{d\theta}\right|_{\theta=\theta_n^0}\Delta\theta_n+
\mathcal{O}(\Delta\theta_n^2)=-\frac{64}{243}\frac{W_n+1}{W_n^2}\widetilde{D}_{OL}\Delta\theta_n+
\mathcal{O}(\Delta\theta_n^2),
\end{equation}
where $\Delta\theta_n=\theta_n-\theta_n^0$. The relative error for $\theta_0^1$ is $|\Delta\theta_1|/\theta_1^0$. Using (\ref{***}) we find that
\[
\frac{|\Delta\theta_1|}{\theta_1^0}=\left(\frac{243 W^2_1\varphi_1}{64\widetilde{D}_{OL}\theta_1^0(W_1+1)}\right)\frac{|\Delta\varphi_1|}{\varphi_1}. 
\]
In the case of the first image we can take $\varphi_1=2\pi$. Moreover, we also have $|\Delta\varphi_1|/\varphi_1\approx 0.001$ for $\rho_0=1.55$. By means of (\ref{box}) with $n=1$ we find that the corresponding impact parameter is
\[
\widetilde{b}_1=\theta_1^0\widetilde{D}_{OL}=\frac{3\sqrt{3}}{2}+\frac{243}{64}W_1^2 
\]
and employing (\ref{punto}) we get that the closest approach distance for the first image is
\[
\rho_{0,1}=\frac{3}{2}-\frac{9}{8}W_1\approx 1.5339 
\]
instead of $1.545$ as given in \cite{Bozza1}. In Table~I we listed the impact paramaters, and the distances of closest approach for $n=1,\cdots,4$. Finally, we obtain for the relative error for $\theta_0^1$
\[
\frac{|\Delta\theta_1|}{\theta_1^0}=\frac{486\pi W_1^2}{(W_1+1)(96\sqrt{3}+243 W_1^2)}\frac{|\Delta\varphi_1|}{\varphi_1}\approx 8.6\cdot 10^{-6}
\]
instead of $8\cdot 10^{-5}$ as given in \cite{Bozza1}. Furthermore, equation (\ref{***}) can be replaced into the lens equation (\ref{II}) to give
\begin{equation}\label{4}
\beta=\theta_n^0+\left(1+\frac{64}{243}\frac{\widetilde{D}_{LS}\widetilde{D}_{OL}}{\widetilde{D}_{OS}}\frac{W_n+1}{W_n^2}\right)\Delta\theta_n, 
\end{equation}
which represents the position of the $n$-th image. Since in general $\widetilde{D}_{LS}\widetilde{D}_{OL}/\widetilde{D}_{OS}\gg 1$ and 
$(64/243)(W_n+1)W_n^{-2}\approx 280.82$, the second term in the 
bracket in (\ref{4}) is much bigger than one and therefore, we can approximate (\ref{4}) as follows
\begin{equation}\label{5}
\beta=\theta_n^0+\frac{64}{243}\frac{\widetilde{D}_{LS}\widetilde{D}_{OL}}{\widetilde{D}_{OS}}\frac{W_n+1}{W_n^2}\Delta\theta_n.
\end{equation}
Finally, from (\ref{5}) we obtain that the position of the $n$-th image is
\begin{equation}\label{5*}
\theta_n=\theta_n^0+\frac{243}{64}\frac{(\beta-\theta_0^n)\widetilde{D}_{OS}}{\widetilde{D}_{LS}\widetilde{D}_{OL}}\frac{W_n^2}{W_n+1}. 
\end{equation}
From (\ref{4}) we have $d\beta/d\theta>0$. This implies that $\mu_r>0$ in (\ref{tr}) and therefore there are no radial critical curves. On the other hand, $\mu_t$ becomes singular when $\beta=0$. Hence, the only critical curves are of tangential nature and since we already solved the lens equation, it is sufficient to set $\beta=0$ in (\ref{5*}) to obtain 
\[
\theta_{n,cr}=\left(1-\frac{243}{64}\frac{\widetilde{D}_{OS}}{\widetilde{D}_{LS}\widetilde{D}_{OL}}\frac{W_n^2}{W_n+1}\right)\theta_n^0. 
\]
The magnification of the $n$-th image is given by the formula \cite{Bozza1} 
\[
\mu_n=\frac{\theta_0^n}{\beta\left.\frac{d\beta}{d\theta}\right|_{\theta=\theta_0^n}}.
\]
Since
\[
\left.\frac{d\beta}{d\theta}\right|_{\theta=\theta_0^n}=1+\frac{64}{243}\frac{\widetilde{D}_{LS}\widetilde{D}_{OL}}{\widetilde{D}_{OS}}\frac{W_n+1}{W_n^2},
\]
where the second term is much grater than one, the magnification of the $n$-th image is
\begin{equation}\label{magg}
\mu_n=\frac{\widetilde{D}_{OS}B_n}{\beta\widetilde{D}^2_{OL}\widetilde{D}_{LS}},\quad
B_n=\frac{243}{4096}\frac{W^2_n(96\sqrt{3}+243 W^2_n)}{W_n+1}.
\end{equation}
and it decreases very quickly because $B_1=9.2\cdot 10^{-3}$, $B_2=1.5\cdot 10^{-5}$, $B_3=2.9\cdot 10^{-8}$. Hence, the luminosity of the first image will dominate over all others. Finally the total magnification 
is \cite{Bozza1}
\begin{equation}\label{maggt}
\mu_{tot}=2\sum_{n=1}^{\infty}\mu_n=\frac{\widetilde{D}_{OS}}{\beta\widetilde{D}^2_{OL}\widetilde{D}_{LS}}\sum_{n=1}^\infty B_n. 
\end{equation}
Since $B_4$ is of the order $10^{-11}$, the series above is rapidly convergent and a good approximation for the total magnification is given by
\[
\mu_{tot}=\frac{\widetilde{D}_{OS}B}{\beta\widetilde{D}^2_{OL}\widetilde{D}_{LS}},\quad
B=B_1+B_2+B_3. 
\]
Our formulae (\ref{magg}) and (\ref{maggt}) generalize equations $(24)$ and $(27)$ in \cite{Bozza1}. 
The amplification of each of the weak field images is \cite{Bozza1}
\[
\mu_{wfi}=\frac{1}{\beta}\sqrt{\frac{2\widetilde{D}_{LS}}{\widetilde{D}_{OL}\widetilde{D}_{OS}}}.
\]
Then,
\[
\frac{\mu_{wfi}}{\mu_{tot}}=\frac{\sqrt{2}}{B}\left(\frac{\widetilde{D}_{LS}\widetilde{D}_{OL}}{\widetilde{D}_{OS}}\right)^{3/2}
\]
with $B\approx 0.0092$ instead of $0.017$ as in \cite{Bozza1}. This implies that relativistic images are extremely faint in comparison to the weak 
field images. Moreover, we also find that the observables $\theta^0_\infty$ and 
$s=\theta^0_1-\theta^0_\infty$ whose numerical values are given in Table I in \cite{Bozza2} should 
read $16.89$ $\mu\mbox{arcsec}$ and $0.02245$ $\mu\mbox{arcsec}$ instead of $16.87$ $\mu\mbox{arcsec}$ and $0.0211$ $\mu\mbox{arcsec}$ in the case of a lens represented by a Schwarzschild black hole with mass $M=2.8\cdot 10^6 M_\odot$ and $D_{OL}=8.5$ Kpc. Last but not least, the sequence of impact parameters can be computed according to the formula
\[
\widetilde{b}_n=\widetilde{b}_{cr}\left(1+\frac{27\sqrt{3}}{32}W^2_n\right). 
\]
The angular size $\theta^0_n$ of the relativistic rings generated by photons deflected by $2\pi,4\pi,6\pi,$ etc will be given by 
$\theta^0_n=\widetilde{b}_n/\widetilde{D}_{OL}$. Note that the above formula generalizes equation $(15)$ in \cite{Kogan}. Employing (\ref{punto}) 
we can compute the corresponding distances of closest approach. In Table I we give the first five values of the impact parameters, the corresponding 
distances of closest approach and the size of the relativistic rings. Finally, from (\ref{Einstein}) we find that the position of the main ring is $\theta_E=1.157987$ arcsec.
\begin{table}[H] 
\caption{Impact parameters, distances of closest approach and positions of the relativistic rings for the black hole believed to be hosted 
by our galaxy ($M=2.8\cdot 10^6 M_\odot$ and $D_{OL}=8.5$ Kpc \cite{ric})} 
\begin{center}
\begin{tabular}{ | l | l | l | l|}
    \hline
    $n$ & $\widetilde{b}_n$ & $\rho_{0,n}$ & $\theta^0_n$ $\mu$arcsec  \\ \hline
    1 & 2.601529850 & 1.533929564  & 16.91272357              \\ \hline
    2 & 2.598082299 & 1.501424471  & 16.89031076              \\ \hline
    3 & 2.598076223 & 1.500061482  & 16.89027125              \\ \hline
    4 & 2.598076211 & 1.500002656  & 16.89027118               \\ \hline
\end{tabular}
\label{table1}
\end{center}
\end{table}

\section{Bending of light in the Schwarzschild - deSitter metric}
The Schwarzschild - deSitter metric describes a static black hole of mass $M$ in a universe with positive cosmological constant $\Lambda$. The 
corresponding line element is given by (\ref{25.3}) with 
\[
B(r)=1-\frac{r_S}{r}-\frac{r^2}{r^2_\Lambda},\quad A(r)=\frac{1}{B(r)},\quad C(r)=1, \quad r_S=2M,\quad r_\Lambda=\sqrt{\frac{3}{\Lambda}}. 
\]
The numerical value of the cosmological constant is extremely small and can be taken to be $\Lambda\approx 10^{-56}$ cm$^{-2}$  
according to \cite{Krauss} even though astrophysical tests such as the perihelion precession of planets in our solar system and other tests based on 
the large scale geometry of the universe suggest an upper bound for the cosmological constant given by $\Lambda\lesssim 10^{-42}~\mbox{m}^{-2}$ 
\cite{Cardona,Jetzer,Sereno1,Iorio}.  
With the inclusion of $\Lambda$ gravity becomes essentially
a two scale theory. It might appear that whereas $\Lambda$
governs only the cosmological aspects, the only constant
entering local gravity effects is the Newtonian constant.
However, in certain circumstances a combination of the
two scales also appears. For instance, for massive
particles the effective potential (\ref{effpot}) for the
Schwarzschild-de Sitter metric develops a local
maximum of the astrophyscal order of magnitude $(r_Sr_{\Lambda})^{1/3}$
signifying the largest radius of a bound state \cite{m1}.
The valid question is if lensing in the Scharzschild-de Sitter
metric gives us also some surprises.

Following \cite{SH} we rescale the time and radial coordinates as $\widetilde{t}=t/r_S$ and $\rho=r/r_S$. Then, the 
metric function $B$ can be rewritten in terms of only one dimensionless parameter $y$ as  
\[
B(\rho)=1-\frac{1}{\rho}-y\rho^2,\quad y=\frac{r_S^2}{r_\Lambda^2}=\frac{4}{3}M^2\Lambda. 
\]
Note that the relation between our parameter $y$ and the corresponding one in \cite{SH} reads $y_s=y/4$. Instead of a single event horizon as in the 
Schwarzschild metric there are different possibilities for the Schwarzschild - deSitter metric
\begin{enumerate}
\item 
two distinct horizons for $0<y<4/27$ located at 
\[
\rho_h=\frac{2}{\sqrt{3y}}\cos{\frac{\pi+\psi}{3}},\quad\rho_c=\frac{2}{\sqrt{3y}}\cos{\frac{\pi-\psi}{3}},\quad
\psi=\cos^{-1}\left(\frac{3}{2}\sqrt{3y}\right).
\]
where $\rho_h$ and $\rho_c$ denote the event and cosmological horizon, respectively. Note that the condition $0<y<4/27$ ensures that 
$\cos\psi\in(0,1)$. Moreover, we also have a negative root located at $\rho_{-}=-(2/\sqrt{3y})\cos(\psi/3)$. Finally, if we expand the horizons 
with respect to the parameter $y$ as
\[
\rho_h=1+y+\mathcal{O}(y^2),\quad \rho_c=\frac{1}{\sqrt{y}}-\frac{1}{2}-\frac{3}{8}\sqrt{y}-\frac{y}{2}-\frac{105}{128}y^{3/2}+\mathcal{O}(y^2),
\]
we can verify that the formulae for the event and cosmological horizons predict correctly that $\rho_h\to 1$ and $\rho_c\to+\infty$ for $y\to 0^{+}$ 
as it should be in the Schwarzschild case.
\item
If $y>4/27$, there is only one real root of the equation $B(\rho)=0$ and the space-time describes a naked singularity located at
\[
\rho_n=-\frac{2}{\sqrt{3y}}\cosh{\frac{\psi}{3}},\quad\psi=\cosh^{-1}\left(\frac{3}{2}\sqrt{3y}\right). 
\]
\item
In the case $y=4/27$ the event and cosmological horizons coincide at $\rho_h=3/2=\rho_c$ and there is also a negative root at $\widetilde{\rho}=-3$.
\end{enumerate}
We will analyze gravitational lensing for the case $0<y<4/27$. By means of equation $(3)$ in \cite{Bozza2} the photon sphere is found to be 
again at $\rho_\gamma=3/2$ as it is the case for the Schwarzschild metric. According to \cite{SH} the critical parameter of the photon circular 
orbit depends on $y$ and is given by
\[
\widetilde{b}_c(y)=\frac{3\sqrt{3}}{\sqrt{1-\frac{27}{4}y}}. 
\]
Note that expanding $\widetilde{b}_c$ around $y$ we have $\widetilde{b}_c(y)=3\sqrt{3}+(81\sqrt{3}/8)y+\mathcal{O}(y^2)$ and in the limit $y\to 0$ it 
reproduces correctly the critical values of the impact parameter in the Schwarzschild case that distinguishes photons which fall into the black hole 
from those escaping at infinity. Since the Schwarzschild - deSitter manifold is spherically symmetric, there is no loss in generality if we suppose 
that the positions of the light source and that of the observer belong to the equatorial plane. Using formula (\ref{25.20}) yields
\[
\frac{d\varphi}{d\rho}=\pm\frac{1}{\rho^2}\frac{1}{\sqrt{\frac{1}{\widetilde{b}^2}-\frac{1}{\rho^2}
\left(1-\frac{1}{\rho}-y\rho^2\right)}}
\]
and with the help of the transformation $u=1/\rho$, we find that the photon motion will be governed by the equation
\begin{equation}\label{binet}
\frac{d\varphi}{du}=\mp\frac{1}{\sqrt{u^3-u^2+a}},\quad a=y+\frac{1}{\widetilde{b}^2}. 
\end{equation}
Note that the above differential equation agrees with $(10)$ in \cite{Arakida} and it clearly contains the cosmological constant. The dependence on $\Lambda$ can be removed if we use (\ref{binet}) to derive a second order nonlinear differential equation for $u=u(\varphi)$ as in\cite{Islam}. Furthermore, p hotons with $\widetilde{b}<\widetilde{b}_c$ are doomed to crash into the central singularity, while those characterized by 
$\widetilde{b}>\widetilde{b}_c$ will be able to escape the gravitational pull of the black hole and they will eventually reach the cosmological 
horizon. We are interested in the latter case. If we look back at the term under the square root in (\ref{binet}), we realize that we need to 
introduce a motion reality condition represented by  $u^3-u^2+a\geq 0$. Turning points will be represented by the roots of the associated cubic 
equation. To study the existence of these points it is more efficient to switch back to the radial variable $\rho$ and consider the cubic 
equation $a\rho^3-\rho+1=0$ which is in principle the same equation we would obtain by setting $B(\rho)=0$ with the parameter $y$ replaced by $a$. 
Since we already analyzed the latter equation, we can immediately conclude that we have the following three cases
\begin{enumerate}
\item 
two turning points for $0<a<4/27$ at 
\[
\rho_{1}=\frac{2}{\sqrt{3a}}\cos{\frac{\pi+\beta}{3}},\quad\rho_2=\frac{2}{\sqrt{3a}}\cos{\frac{\pi-\beta}{3}},\quad
\beta=\cos^{-1}\left(\frac{3}{2}\sqrt{3a}\right)
\]
such that $\rho_h<\rho_1<\rho_\gamma<\rho_2<\rho_c$. Moreover, the cubic $a\rho^3-\rho+1$ will be negative on the interval $(\rho_\gamma,\rho_1)$ and 
positive on $(\rho_2,\rho_c)$.
\item
If $a>4/27$, there is only one turning point at
\[
\rho_t=-\frac{2}{\sqrt{3a}}\cosh{\frac{\beta}{3}},\quad\beta=\cosh^{-1}\left(\frac{3}{2}\sqrt{3a}\right)
\]
and since $\rho_t<0$ the cubic will be positive on the interval $(\rho_h,\rho_c)$.
\item
In the case $a=4/27$ the turning points found in $1.$ coalesce into a single turning point at $\rho_t=\rho_\gamma=3/2$ and the cubic 
will be positive on the interval $(\rho_\gamma,\rho_c)$.
\end{enumerate}
It is interesting to observe that the conditions $\widetilde{b}>\widetilde{b}_c$ and $0<y<4/27$ roule out the possibility that $a\geq 4/27$ as it can 
be seen from the following simple estimate
\[
a=y+\frac{1}{\widetilde{b}^2}<y+\frac{1}{\widetilde{b}^2_c}=\frac{3}{4}y+\frac{1}{27}<\frac{4}{27}. 
\]
Hence, we must have $0<y<a<4/27$. Let $\rho_0$ denote the distance of closest approach. According to the above discussion we must take $\rho_0$ in 
the interval $(\rho_2,\rho_c)$. By $\rho_{b}$ we will denote the position of the observer. Then, from (\ref{25.20}) we find that 
\begin{equation}\label{soluza}
\varphi(\rho_0)=\int_{\rho_0}^{\rho_b}\frac{d\rho}{\rho}\sqrt{A(\rho)}\left[\left(\frac{\rho}{\rho_0}\right)^2 \frac{B(\rho_0)}{B(\rho)}-1\right]^{-1/2}. 
\end{equation}
The main difference with the Schwarzschild case is that the observer cannot be positioned asymptotically in a region where we can assume that 
the space-time is described by the Minkowski metric. For this reason we will suppose that the deflection angle is given by the formula 
\begin{equation}\label{formula}
\Delta\varphi(\rho_0)=\kappa_1 I(\rho_0)+\kappa_2 
\end{equation}
where $I(\rho_0)$ denotes the integral in (\ref{soluza}) and $\kappa_1$ and $\kappa_2$ are two scalars to be determined in such a way that the weak 
field approximation of (\ref{formula}) reproduces the weak field approximation for the Schwarzschild case in the limit $y\to 0$. Let $x=\rho/\rho_0$. 
Then, we have
\[
I(\rho_0)=\int_1^{x_b}\frac{dx}{x\sqrt{x^2 B(\rho_0)-B(\rho_0 x)}}. 
\]
Let $\alpha=1/\rho_0$. The term under the square root in the above expression becomes 
\[
x^2 B(\rho_0)-B(\rho_0 x)=(1-\alpha)x^2+\frac{\alpha}{x}-1
\]
and it does not depend on the cosmological constant $\Lambda$. This is not surprising since it is well known that $\Lambda$ has an influence over the orbits of massive particles but it can be made to disappear 
from the coordinate orbital equation when photons are considered \cite{Islam}. At this point the integral $I$ will depend only on the parameter $\alpha$ and will be given by
\[
I(\alpha)=\int_1^{x_b}\frac{dx}{x}\left[(1-\alpha)x^2+\frac{\alpha}{x}-1\right]^{-1/2}.
\]
Letting $\rho_0\gg 1$ corresponding to the condition $\alpha\ll 1$ we can expand the integral $I$ in powers of the small parameter $\alpha$ 
according to
\begin{equation}\label{Ialpha}
I(\alpha)=I(0)+I^{'}(0)\alpha+\frac{I^{''}(0)}{2}\alpha^2+\frac{I^{'''}(0)}{3!}\alpha^3+\mathcal{O}(\alpha^4),
\end{equation}
where the prime denotes differentiation with respect to $\alpha$. Note that we can differentiate under the integral since the integrand is 
continuous on the interval $(1,x_b)$. The coefficients in the above expansion have been computed with the software 
Maple 14 and they are given by the following formulae
\begin{eqnarray*}
I(0)&=&\frac{\pi}{2}-f_1(x_b),\quad f_1(x_b)=\arctan{\left(\frac{1}{\sqrt{x^2_b-1}}\right)},\\
I^{'}(0)&=&\frac{2x^2_b-x_b-1}{2x_b\sqrt{x^2_b-1}},\\
I^{''}(0)&=&\frac{15x^2_b(x_b+1)^2[\pi-2f_1(x_b)]-2\sqrt{x^2_b-1}f_2(x_b)}{16x^2_b(x_b+1)^2},\quad 
f_2(x_b)=8x^3_b+7x^2_b-6x_b-3,\\
I^{'''}(0)&=&-\frac{45x^3_b(x_b+1)^3[\pi-2f_1(x_b)]-2\sqrt{x^2_b-1}f_3(x_b)}{16x^3_b(x_b+1)^3},\quad
f_3(x_b)=122x^5_b+306x^4_b+247x_b^3+70x^2_b+15 x_b+5.
\end{eqnarray*}
Moreover, for $x_b\gg 1$ the above quantities can be expanded as follows 
\begin{equation}\label{as1}
I(0)=\frac{\pi}{2}-\frac{1}{x_b}+\mathcal{O}(x_b^{-2}),\quad
I^{'}(0)=1-\frac{1}{2x_b}+\mathcal{O}(x_b^{-2}),
\end{equation}
\begin{equation}\label{as2}
I^{''}(0)=\frac{15}{16}\pi-1-\frac{3}{4x_b}+\mathcal{O}(x_b^{-2}),\quad 
I^{'''}(0)=\frac{61}{4}-\frac{45}{16}\pi-\frac{15}{8x_b}+\mathcal{O}(x_b^{-2}).
\end{equation}
Hence, in the limit $x_b\to\infty$ corresponding to $y\to 0$ we find that the deflection angle is given by
\[
\Delta\varphi(\rho_0)=\frac{\pi}{2}\kappa_1+\kappa_2+\kappa_1\rho_0^{-1}+\frac{\kappa_1}{2}\left(\frac{15}{16}\pi-1\right)\rho_0^{-2}+
\frac{\kappa_1}{6}\left(\frac{61}{4}-\frac{45}{16}\pi\right)\rho_0^{-3}+\mathcal{O}(\rho_0^{-4}). 
\]
Comparison of the above expression with (\ref{wfl}) gives $\kappa_1=2$ and $\kappa_2=-\pi$. Letting 
$x_b$ to approach the cosmological horizon at $x_c$ and employing (\ref{as1}) and (\ref{as2}) it is not difficult to verify that the deflection angle can be written at the third order in the parameter $\alpha$ and at the first order in $1/x_c$ as
\[
\Delta\varphi(\alpha)=-\frac{2}{x_c}+\mathcal{O}\left(\frac{1}{x^2_c}\right)+\left[2-\frac{1}{x_c}+\mathcal{O}\left(\frac{1}{x^2_c}\right)\right]\alpha+
\left[\frac{15}{16}\pi-1-\frac{3}{4x_c}+\mathcal{O}\left(\frac{1}{x^2_c}\right)\right]\alpha^2+
\]
\[
+\left[\frac{1}{3}\left(\frac{61}{4}-\frac{45}{16}\pi\right)-\frac{5}{8x_c}+\mathcal{O}\left(\frac{1}{x^2_c}\right)\right]\alpha^3+
\mathcal{O}(\alpha^4).
\]
Finally, expanding the cosmological horizon in powers of $\Lambda$ we find that the deflection angle as a function of the distance of closest approach for an observer located asymptotically at the cosmological horizon can be approximated by the following formula
\[
\Delta\varphi(r_0)=-\frac{2}{\sqrt{3}}r_0\sqrt{\Lambda}+\left(2-\frac{r_0\sqrt{\Lambda}}{\sqrt{3}}\right)\frac{2M}{r_0}+\left[\left(\frac{15}{16}\pi-1\right)-\frac{\sqrt{3}}{4}r_0\sqrt{\Lambda}\right]\left(\frac{2M}{r_0}\right)^2+
\]
\begin{equation}\label{def_SdS}
+\left[\frac{1}{3}\left(\frac{61}{4}\pi-\frac{45}{16}\pi\right)-\frac{5}{8\sqrt{3}}r_0\sqrt{\Lambda}\right]\left(\frac{2M}{r_0}\right)^3+\cdots
\end{equation}
Hence, we agree with \cite{Rindler_1} that weak gravitational lensing in the Schwarzschild-deSitter metric will depend on the cosmological constant 
$\Lambda$. It is interesting to observe that equation $(18)$ in \cite{Rindler_2} in the limit $\Lambda\to 0$ fails to reproduce correctly the 
term going together with $(2M/r_0)^2$ in the weak field limit of the Schwarzschild metric, while our (\ref{def_SdS}) matches the corresponding formula 
in the Schwarzschild case even at the order $(2M/r_0)^3$. \cite{Sereno} considered the light orbital equation with the source located at 
$(r_s,\varphi_s)$ while the observer is positioned at $(r_b,\varphi_b)$ with $\varphi_b=0$ and derived an expression for $\varphi_s$ in powers of 
$1/b$, $1/r_s$, and $1/r_b$. Concerning strong gravitational lensing in a Schwarzschild-deSitter manifold we will first solve exactly the integral 
in (\ref{Ialpha}) in terms of an incomplete elliptic integral of the first kind and then apply an asymptotic formula derived by \cite{carlson} 
in the case when the sine of the modular angle and the elliptic modulus both tends to one. To this purpose we write the deflection angle as 
$\Delta\varphi(\alpha)=2I(\alpha)-\pi$ where the integral $I(\alpha)$ is given by (\ref{Ialpha}) and introduce the coordinate transformation 
$u=1/x$ so that we obtain
\[
\Delta\varphi(\alpha)=2\int_{1/x_b}^1\frac{du}{\sqrt{\alpha u^3-u^2+1-\alpha}}-\pi. 
\]
The cubic under the square root in the above expression as zeroes at
\[
u_0=1,\quad u_1=\frac{1-\alpha-\sqrt{1+2\alpha-3\alpha^2}}{2\alpha},\quad 
u_2=\frac{1-\alpha+\sqrt{1+2\alpha-3\alpha^2}}{2\alpha}. 
\]
Since $\rho_0>\rho_2>\rho_\gamma$, then $\alpha<2/3$ and we can order the roots as $u_2>u_0>0>u_1$. Using formula $3.131.4$ in \cite{Gradsh} we can 
express the deflection angle in terms of an incomplete elliptic integral of the first kind as follows
\[
\Delta\varphi(\alpha)=-\pi+\frac{4F(\phi_1,\kappa)}{\sqrt{\alpha(u_2-u_1)}},\quad
\phi_1=\sin^{-1}{\sqrt{\frac{(u_2-u_1)(1-1/x_b)}{(1-u_1)(u_2-1/x_b)}}},\quad
\kappa=\sqrt{\frac{1-u_1}{u_2-u_1}}.
\]
Note that $\kappa$ can be expanded around the photon sphere as in (\ref{kexp}) while the modular angle admits the Taylor expansion
\[
\sin{\phi_1}=1-\frac{4}{9}\frac{3+\rho_b}{2\rho_b-3}(\rho_0-\rho_\gamma)+\mathcal{O}(\rho_0-\rho_\gamma), 
\]
which agrees with the corresponding expansion in (\ref{kexp}) when $\rho_b\to\rho_c$ and $\Lambda\to 0$. Since both $\kappa$ and $\sin{\phi_1}$ 
tends to $1$ as $\rho_0\to\rho_\gamma$, we can apply the asymptotic expansion (\ref{expansion}) for the incomplete elliptic integral 
of the first kind developed by \cite{carlson} and the deflection angle can be approximated as
\[
\Delta\varphi(\rho_0)=-\pi+\ln{\frac{144(2\rho_b-3)}{[3+4\rho_b+2\sqrt{3\rho_b(3+\rho_b)}]^2}}-2\ln{\left(\frac{\rho_0}{\rho_\gamma}-1\right)}
+\mathcal{O}(\rho_0-\rho_\gamma). 
\]
The above formula reduces correctly to the corresponding one derived in the previous section for the Schwarzschild metric when $\rho_b\to\rho_c$ 
and $\Lambda\to 0$. Finally, if $\rho_b\to\rho_c$ in the above expression and we make an expansion around $\Lambda=0$ we obtain
\[
\Delta\varphi(r_0)=-\pi+\ln{144(7-4\sqrt{3})}-\frac{2}{3}(9-2\sqrt{3})M\sqrt{\Lambda}+\mathcal{O}(M^2\Lambda)-2\ln{\left(\frac{r_0}{r_\gamma}-1\right)}
+\mathcal{O}(r_0-r_\gamma). 
\]

\section{Light deflection in the Janis - Newman - Winicour metric}
The Janis - Newman - Winicour metric is the most general spherically symmetric static and asymptotically flat solution of Einstein's field 
equations coupled to a massless scalar field and is given by \cite{Janis}
\begin{equation}\label{WPM}
ds^2=\left(1-\frac{\mu}{r}\right)^\gamma dt^2-\left(1-\frac{\mu}{r}\right)^{-\gamma}dr^2-r^2\left(1-\frac{\mu}{r}\right)^{1-\gamma}
(d\vartheta^2+\sin^2{\vartheta}d\varphi^2)
\end{equation}
with $\gamma=M/\sqrt{M^2+q^2}$ and $\mu=2\sqrt{M^2+q^2}$ where $M$ is the total mass and $q$ is the strength of the scalar field also called the 
``scalar charge''.  Note that $\gamma\leq 1$. For $\gamma=1$ the JNW metric reduces to the Schwarzschild solution and the scalar field vanishes. 
The metric has been re-discovered by Wyman \cite{wyman} and his solution was shown to be
equivalent to the JNW metric in \cite{Vyb}. The metric is not only interesting
from the point of view of an example of a naked singularity, but has served as a model for the
supermassive galactic center in \cite{center}.

In order to study gravitational lensing it is convenient to rescale the time and radial coordinates as 
$\widetilde{t}=t/\mu$ and $\rho=r/\mu$. Then, the above metric can be cast into the form 
\[
d\widetilde{s}^2=\frac{ds^2}{\mu^2}=\left(1-\frac{1}{\rho}\right)^\gamma d\widetilde{t}^2
-\left(1-\frac{1}{\rho}\right)^{-\gamma}d\rho^2-\rho^2\left(1-\frac{1}{\rho}\right)^{1-\gamma}
(d\vartheta^2+\sin^2{\vartheta}d\varphi^2),\quad\gamma=\frac{r_S}{\mu}.
\]
\cite{virba} and \cite{amore} constructed weak field approximations of the deflection angle up to the second order. We point out that formula $(51)$ 
in \cite{amore} reproduces correctly the first order term of the weak field limit of the Schwarzschild metric but the second order term fails to do 
so when $\nu\to 0$ in the aforementioned formula. Before going into the details of light bending in this metric, let us have a closer look at the 
nature of the naked singularity of this gravitational background for $\rho=1$. The existence of the latter is indicated by the Kretschmann invariant 
given by
\[
K=R^{abcd}R_{abcd}=\frac{\mu^2(r-\mu)^{2\gamma}}{4r^{2\gamma+4}(r-\mu)^4}f_\gamma(r),\quad 
f_\gamma(r)=48\gamma^2 r^2-16\mu\gamma(\gamma+1)(2\gamma+1)r+\mu^2(\gamma+1)^2(7\gamma^2+2\gamma+3).
\]
At the same time it is clear that for certain choices of $\gamma$ the coordinate system used to write the line element (\ref{WPM}) does not represent 
the full maximal atlas. Indeed, it is possible to find a coordinate system where the metric elements are nonsingular and the naked singularity 
manifests itself through the noninvertibility of the metric. As an example let us choose $\gamma=1/3$. The null radial geodesic gives rise to the 
definition of the tortoise coordinate $r_{*}$ via the first order differential equation
\[
\frac{dr_{*}}{dr}=\frac{1}{\sqrt[3]{1-\frac{\mu}{r}}}. 
\]
Using the integral representation of the hypergeometric function (see $15.3.1$ in \cite{abra}) one obtains
\[
r_{*}=-\frac{3}{4}r\sqrt[3]{\frac{r}{\mu}} {}_2 F_1\left(\frac{1}{3},\frac{4}{3};\frac{7}{3};\frac{r}{\mu}\right).
\]
The analog of the Eddington-Finkelstein coordinates for this metric is now $\widetilde{u}=t+r_{*}$ and $\widetilde{v}=t-r_{*}$. In these coordinates 
the line element takes now the form
\[
ds^2=\sqrt[3]{1-\frac{\mu}{r}}d\widetilde{u}^2-2d\widetilde{u}dr-C(r)(d\vartheta^2+\sin^2{\vartheta}d\varphi^2)
\]
from which one can see that the presence of the naked singularity is not obvious. On the other hand, the determinant of the metric is
\[
\mbox{det}g=-r^4\left(1-\frac{\mu}{r}\right)^{4/3}\sin^2{\theta} 
\]
which makes the metric not to be invertible at $r=\mu$. This illustrative example shows the subtle nature of the naked singularity. Here, by means 
of an alternative method we offer a formula extending the latter expansions up to the fifth order and reducing correctly to the weak 
field approximation (\ref{wfl}) in the limit of a vanishing ``scalar charge''.  With the help of $(7)$ in \cite{Bozza2} the integral giving the 
deflection angle can be written as $\Delta\varphi(\rho_0)=I(\rho_0)-\pi$ where
\[
I(\rho_0)=2\int_{\rho_0}^\infty\frac{d\rho}{\rho}\sqrt{\frac{A(\rho)}{C(\rho)}}
\left[\left(\frac{\rho}{\rho_0}\right)^2\frac{C(\rho)B(\rho_0)}{C(\rho_0)B(\rho)}-1\right]^{-1/2},
\quad C(\rho)=\left(1-\frac{1}{\rho}\right)^{1-\gamma}.
\]
Letting $\alpha=1/\rho_0$ and introducing the change of variable $u=(\alpha\rho)^{-1}$ we obtain
\[
I(\alpha)=2\int_0^1\left[(1-\alpha u)^{2-2\gamma}(1-\alpha)^{2\gamma-1}-u^2(1-\alpha u)\right]^{-1/2}.
\]
Expanding the integrand in $I(\alpha)$ around $\alpha=0$ we obtain after a tedious computation
\[
I(\alpha)=\pi+2\gamma\alpha+c_1\alpha^2+\cdots+c_4\alpha^5+\mathcal{O}(\alpha^6) 
\]
with
\begin{eqnarray*}
c_1&=&(\pi-2)\gamma^2+\gamma-\frac{\pi}{16},\\
c_2&=&(7-2\pi)\gamma^3+(\pi-2)\gamma^2+\frac{1}{4}\left(\frac{1}{3}+\frac{\pi}{2}\right)\gamma-\frac{\pi}{16},\\ 
c_3&=&\left(6\pi-\frac{55}{3}\right)\gamma^4+\left(\frac{21}{2}-3\pi\right)\gamma^3+\frac{1+3\pi}{12}\gamma^2+\frac{3}{8}\left(\frac{\pi}{2}-1\right)\gamma-\frac{55}{1024}\pi,\\
c_4&=&\left(\frac{1975}{36}-\frac{52}{3}\pi\right)\gamma^5+\left(12\pi-\frac{110}{3}\right)\gamma^4+\left(\frac{323}{72}-\frac{13}{12}\pi\right)\gamma^3+\left(\frac{13}{6}-\frac{\pi}{2}\right)\gamma^2+\left(\frac{149}{768}\pi-\frac{1513}{2880}\right)\gamma-\frac{53}{512}\pi. 
\end{eqnarray*}
Hence, the weak field limit of the deflection angle reads
\[
\Delta\varphi(\rho_0)=\frac{2\gamma}{\rho_0}+\frac{c_1}{\rho_0^2}+\cdots\frac{c_4}{\rho_0^5}+
\mathcal{O}\left(\frac{1}{\rho_0^6}\right).
\]
The above expansion generalizes the weak field approximations $(24)$ in \cite{virba} and $(51)$ in \cite{amore}. Moreover, it reproduces the weak field approximation (\ref{wfl}) for a Schwarzschild manifold in the limit $\gamma\to 1$. For $\gamma=1/2$ the deflection angle can be computed analytically in terms of the complete elliptic integral of the first kind $K$ and the incomplete 
elliptic integral of the first kind $F$ as
\[
\Delta\varphi(\rho_0)=-\pi+2\sqrt{\frac{2}{\rho_0+1}}
\left[K\left(\sqrt{\frac{2}{\rho_0+1}}\right)-F\left(\frac{\sqrt{2}}{2},\frac{2}{\rho_0+1}\right)\right].
\]
The above formula is new since we could not find any similar result in the existing literature concerning gravitational lensing in the JNW metric. 
Concerning a numerical analysis of gravitational lensing for the case $0<\gamma<1/2$ we refer to \cite{ellis} whereas the study of the strong 
gravitational lensing can be found in \cite{Bozza2}. The $\gamma=1/2$ is indeed a special case of the JNW metric. We are interested in the region 
$\rho>1$ assuming that the geodesics cannot be continued through the naked singularity. Then, the function $\widetilde{V}$ defined in (\ref{vtilde}) 
displays a maximum (unstable circular orbit) at $\rho_m=\gamma+1/2$. In the very principle, light can be trapped now between the hump of 
$\widetilde{V}$ and the line $\rho=1$. For $\gamma<1/2$ the function $\widetilde{V}$ is a smoothly decreasing function between one and infinity 
whereas for $\gamma=1/2$ we have $\widetilde{V}(\rho)=1/\rho^2$.

\begin{figure}\label{JNW}
\includegraphics[scale=0.5]{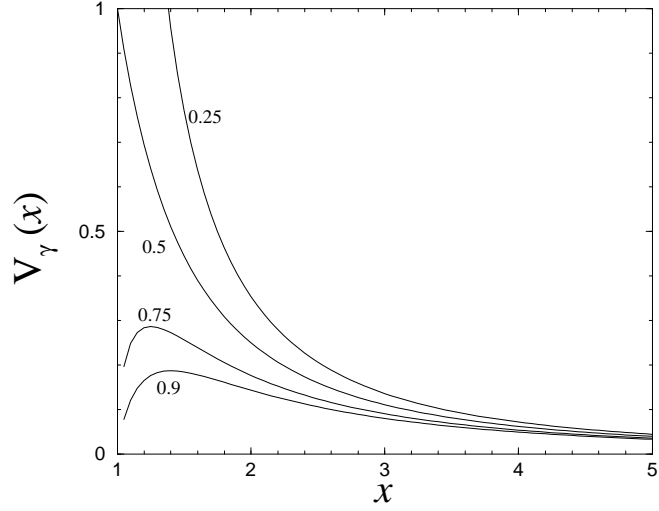}
\caption{\label{fig3.2}
Behaviour of the rescaled ``photon potential'' proportional $\widetilde{V}$ for the JNW
metric (naked singularity) for different values of the parameter $\gamma$. As long as
$\gamma < 1/2$ the metric displays a local maximum at the rescaled position bigger than 1.
Since the non-removable singular point at $x=1$ cannot be crossed light can be trapped, 
in principle, between the singular point and the hump.
}
\end{figure}

\section{Light deflection in a spacetime with galactic dark matter halos}
One of the pressing problems of astrophysics is to explain
the rotational curves of galaxies. The post popular explanation
is to postulate non-baryonic neutral matter (Dark Matter) \cite{DM}.
Another explanation prefers to modify the gravity itself \cite{Mond}.
In both cases, an interesting task is to extend the non-relativistic
theories within the framework of General Relativity. In the
case of DM this means to use existing empirical density profiles
to construct a DM halo metric.

The line element associated to a galaxy Dark matter halo 
based on the Navarro-Frenk-White (NFW) \cite {NFW} density profile 
is a metric of the form (\ref{25.3})  
\cite{Matos1}
\begin{eqnarray*}
B(x)&=&\left\{ \begin{array}{ll}
         1+2\Phi_c+\gamma C_0 x^2 & \mbox{if $0\leq x \leq x_0$}\\
         1+2\Phi_c+\gamma A_0-\frac{\gamma}{x}\left(B_0+\ln{\frac{1+x}{1+x_0}}\right) & \mbox{if $x > x_0$}.\end{array} \right.,\\
A(x)&=&\left\{ \begin{array}{ll}
         1+2\gamma C_0 x^2 & \mbox{if $0\leq x \leq x_0$}\\
         1+\frac{\gamma}{x}\left(D_0+\ln{\frac{1+x}{1+x_0}}+\frac{1}{1+x}\right) & \mbox{if $x > x_0$}.\end{array} \right.
\end{eqnarray*}
with
\[
D_0=\frac{x_0^2-3x_0-3}{3(1+x_0)^2},\quad C_0=\frac{1}{6x_0(1+x_0)^2},\quad B_0=\frac{x_0(4x_0+3)}{3(1+x_0)^2},\quad 
A_0=\frac{3x_0+2}{2(1+x_0)^2} 
\]
and
\[
\gamma=2v_0^2,\quad x=\frac{r}{r_g}
\]
where the two free parameters $v_0$ and $r_g$ are the characteristic speed and radius of the galaxy, respectively. In what follows it will be assumed 
that the characteristic speed of the galaxy is small so that $\gamma\ll 1$. Moreover,
\[
\Phi_c\approx -\epsilon\left[1+\frac{2\ln{(1+c_0)}}{c_0}-\frac{1}{1+c_0}\right],\quad\epsilon=\gamma/2
\]
and
\[
c_0=62.1\times\left(\frac{M_{vir}h}{M_\odot}\right)^{-0.06}(1+\epsilon). 
\]
where $M_{vir}$ is the virial mass. According to \cite{Matos1} for $h=0.7$ we have $10^8\lesssim M_{vir}/M_\odot\lesssim 10^{15}$ which implies that 
$6\lesssim c_0\lesssim 30$. Furthermore, for $M_{vir}/M_\odot$ varying in the aforementioned range $x_0$ is of the order $10^{-4}$ as it can be 
evinced from equation $(29)$ in \cite{Matos1}. At this point a comment is in order. If we let $x_0\to 0$, we recover the metric derived 
in \cite{Matos2}. However, there are two conceptual differences. First of all, in \cite{Matos1} there are two regions to be considered and 
moreover, the calculation is valid only for small $\gamma$. 
The case distinction above is necessary because of the singular
nature of the NFW density profile at the origin. To get around this problem
one replaces the inner region by a regular solution \cite{Matos1}.
We have re-calculated the matching conditions and differ
slifgtly with our results of the function $A$ and $B$ from \cite{Matos1}.

In what follows we are interested in the case of weak gravitational lensing and therefore 
we will assume that the distance of closest approach $\widetilde{x_0}\gg 1$. For a generic spherically symmetric spacetime the deflection angle is 
given by the integral \cite{Bozza2}
\begin{equation}\label{form_gen}
\Delta\varphi(\widetilde{x}_0)=-\pi+2\int_{\widetilde{x}_0}^\infty\frac{\sqrt{A(x)}dx}{x\sqrt{\left(\frac{x}{\widetilde{x}_0}\right)^2
\frac{B(\widetilde{x}_0)}{B(x)}-1}}
\end{equation}
To compute the deflection angle in the case of weak gravitational lensing $(x,\widetilde{x}_0\gg 1)$ we follow the method outlined by \cite{Weinberg} and make 
an asymptotic expansion in the small parameters $x^{-1}$ and $\widetilde{x}_0^{-1}$. Taking into account that
\[
B(x)=1+2\Phi_c+\gamma A_0-\gamma\frac{B_0-\ln{(1+x_0)}}{x}-\gamma\frac{\ln{x}}{x}+\mathcal{O}\left(\frac{1}{x^2}\right)
\]
we find that
\begin{eqnarray*}
\left(\frac{x}{\widetilde{x}_0}\right)^2\frac{B(\widetilde{x}_0)}{B(x)}-1&=&
\left(\frac{x}{\widetilde{x}_0}\right)^2\left[1-\gamma\frac{B_0-\ln{(1+x_0)}}{1+2\Phi_c}\frac{x-\widetilde{x}_0}{x\widetilde{x}_0}+
\frac{\gamma}{1+2\Phi_c}\left(\frac{\ln{x}}{x}-\frac{\ln{\widetilde{x}}_0}{\widetilde{x}_0}\right)\cdots\right]-1,\\
&=&\left[\left(\frac{x}{\widetilde{x}_0}\right)^2-1\right]\left[1+\gamma\omega(x,\widetilde{x}_0)+\cdots\right]
\end{eqnarray*}
with
\[
\omega(x,\widetilde{x}_0)=-\frac{B_0-\ln{(1+x_0)}}{1+2\Phi_c}
\frac{x}{\widetilde{x}_0(x+\widetilde{x}_0)}+\frac{1}{1+2\Phi_c}\frac{x(\widetilde{x}_0\ln{x}-x\ln{\widetilde{x}_0})}
{\widetilde{x}_0(x^2-\widetilde{x}_0^2)}\ll 1. 
\]
Moreover,
\[
A(x)=1+\gamma\frac{D_0-\ln{(1+x_0)}}{x}+\gamma\frac{\ln{x}}{x}+\mathcal{O}\left(\frac{1}{x^2}\right)
\]
and the integrand in (\ref{form_gen}) can be approximated as follows
\[
\frac{\sqrt{A(x)}}{x\sqrt{\left(\frac{x}{\widetilde{x}_0}\right)^2\frac{B(\widetilde{x}_0)}{B(x)}-1}}=
\frac{1}{x\sqrt{\left(\frac{x}{\widetilde{x}_0}\right)^2-1}}
\left[1+\gamma\frac{B_0-\ln{(1+x_0)}}{2(1+2\Phi_c)}\frac{x}{\widetilde{x}_0(x+\widetilde{x}_0)}+
\gamma\frac{D_0-\ln{(1+x_0)}}{2x}+\right.
\]
\[
\left.\frac{\gamma}{2}\frac{\ln{x}}{x}-\frac{\gamma}{2(1+2\Phi_c)}\frac{x(\widetilde{x}_0\ln{x}-x\ln{\widetilde{x}_0})}
{\widetilde{x}_0(x^2-\widetilde{x}_0^2)}\right]. 
\]
Let $h(x)=x^{-1}[(x/\widetilde{x}_0)^2-1]^{-1/2}$. By means of the formulae 
\[
\int_{\widetilde{x}_0}^{\infty}h(x)~dx=\frac{\pi}{2},\quad 
\int_{\widetilde{x}_0}^{\infty}\frac{xh(x)}{x+\widetilde{x}_0}~dx=1,\quad
\int_{\widetilde{x}_0}^{\infty}\frac{h(x)}{x}~dx=\frac{1}{\widetilde{x}_0},
\]
\[
\int_{\widetilde{x}_0}^{\infty}h(x)\frac{x(\widetilde{x}_0\ln{x}-x\ln{\widetilde{x}_0})}
{x^2-\widetilde{x}_0^2}~dx=\ln{\frac{2}{\widetilde{x}_0}},\quad
\int_{\widetilde{x}_0}^{\infty}h(x)\frac{\ln{x}}{x}~dx=\frac{\ln{\widetilde{x}_0}-\ln{2}+1}{\widetilde{x}_0}
\]
we find that the deflection angle can be represented  at the order $\gamma$ as follows
\begin{equation}\label{app1}
\Delta\varphi(\widetilde{x}_0)=\gamma\left[1-\frac{2(1+\Phi_c)}{1+2\Phi_c}\ln{2}+D_0-\ln{(1+x_0)}+
\frac{B_0-\ln{(1+x_0)}}{1+2\Phi_c}\right]\frac{1}{\widetilde{x_0}}+\gamma\frac{2(1+\Phi_c)}{1+2\Phi_c}
\frac{\ln{\widetilde{x}_0}}{\widetilde{x}_0}+\cdots. 
\end{equation}
As $\Phi_c\to 0$ which corresponds to $\epsilon\to 0$ and $x_0\to 0$ the deflection angle behaves as $2\gamma\ln{\widetilde{x}_0}
/\widetilde{x_0}$. 
The above results are the first steps to calculate the deflection
angle of Dark Matter Halo within a general relativistic framework.
We leave the lensing and the strong lensing for future projects.

\section{Gravitational lensing in the presence of a holographic screen}
We recall that a holographic screen can be seen as the event horizon of black hole characterized by a mass spectrum bounded from below by a mass 
of the extremal configuration coinciding with the Planck mass. The metric modelling a holographic screen is \cite{P_0}
\[
ds^2=\left(1-\frac{2ML_p^2 r}{r^2+L^2_p}\right)dt^2- \left(1-\frac{2ML_p^2 r}{r^2+L^2_p}\right)^{-1}dr^2-r^2 d\Omega^2,
\]
where the mass of the holographic screen is 
\begin{equation}\label{XI}
M=\frac{r^2_h+L^2_p}{2L^2_p r_h}, 
\end{equation}
$r_h$ is the radius of the screen, and $L_p$ denotes the Planck length. The above line element admits a pair of distinct horizons at
\begin{equation}\label{XII}
r_{\pm}=L^2_p\left(M\pm\sqrt{M^2-M^2_p}\right), 
\end{equation}
whenever $M>M_p$ where $M_p$ is the Planck mass, while for $M=M_p$ the two horizons merge together and we have an extremal black hole. For $M\gg M_p$ we get the usual 
Schwarzschild metric. Inserting (\ref{XI}) into (\ref{XII}) we find that
\[
r_{+}=r_h,\quad r_{-}=\frac{L^2_p}{r_h}. 
\]
In the present case the function $\widetilde{V}$ defined in (\ref{vtilde}) reads
\[
\widetilde{V}(x)=\frac{1}{L^2_p x^2}\left(1-\frac{mx}{x^2+1}\right),\quad x=\frac{r}{L_p},\quad m=\frac{2M}{M_p}.
\]
It is not dificult to verify that for $0\leq m\leq 2$ we have $\widetilde{V}(x)>0$ for any $x>0$ and monotonically decreasing. However, when $m>2$ 
the same function intersects the positive $x$-axis at
\[
{}_1 x_2=\frac{m\pm\sqrt{m^2-4}}{2} 
\]
so that $\widetilde{V}(x)$ is positive on the intervals $(0,x_1)$ and $(x_2,\infty)$ and negative on $(x_1,x_2)$. It possesses a negative minimum at 
$x_m\in(x_1,x_2)$ and a positive maximum at $x_M\in(x_2,\infty)$ where we will have an unstable circular orbit corresponding to the photon sphere. 

\begin{figure}\label{Holo2}
\includegraphics[scale=0.5]{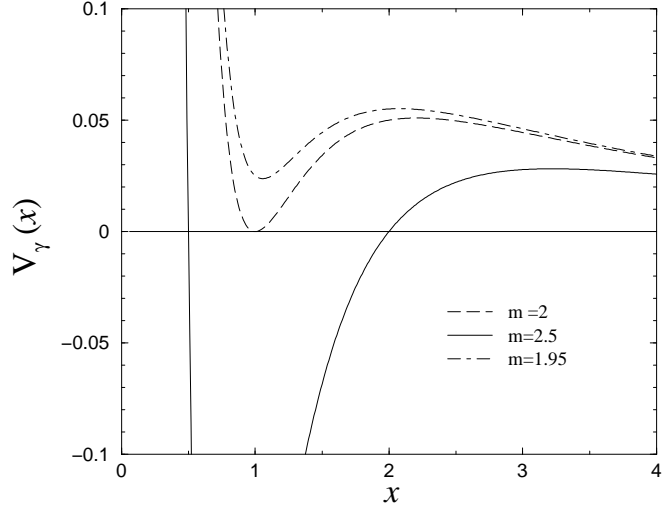}
\caption{\label{fig3.3}
The rescaled ``photon potential'' proportional to $\widetilde{V}$ for the Holographic screen metric.
In case of a naked singularity photons can be trapped in the local minimum.
}
\end{figure}

\subsection{The nonextreme case}
To derive formulae for the deflection angle in the weak and strong regimes we start by considering the nonextreme case $M>M_p$. First of all, we 
rescale $t$ and $r$ by $r_h$. Let $\rho=r/r_h>1$. Then, the original line element can be rewritten as $ds^2=r^2_h d\widetilde{s}^2$ with
\[
d\widetilde{s}^2=B(\rho)d\widetilde{t}^2-\frac{d\rho^2}{B(\rho)}-\rho^2 d\Omega^2,\quad B(\rho)=1-\frac{(1+\lambda)\rho}{\rho^2+\lambda},\quad
\lambda=L^2_p/r^2_h. 
\]
The radius of the photon sphere can be found by solving the equation $B^{'}(\rho)/B(\rho)=2/\rho$ which leads to the problem of finding the roots of the 
quartic polynomial equation
\begin{equation}\label{stella_nc}
p(\rho)=2\rho^4-3(1+\lambda)\rho^3+4\lambda\rho^2-\lambda(1+\lambda)\rho+2\lambda^2=0. 
\end{equation}
For $0<\lambda\ll 1$ we can use a perturbative method to find the position of the photon sphere. In this regard we observe that for $\lambda=0$ the 
unperturbed roots of (\ref{stella_nc}) are $0$ with algebraic multiplicity $3$ and $3/2$. Since the latter has algebraic multiplicity $1$, we can 
set up an expansion $\rho_{3/2}(\lambda)=3/2+x_1\lambda+x_2\lambda^2+x_3\lambda^3+\mathcal{O}(\lambda^4)$. Substituting $\rho_{3/2}(\lambda)$ into 
(\ref{stella_nc}), equating powers of $\lambda$, and solving the corresponding set of equations, we find
\[
\rho_{3/2}(\lambda)=\frac{3}{2}+\frac{7}{18}\lambda+\frac{38}{243}\lambda^2+\frac{1070}{6561}\lambda^4+\mathcal{O}(\lambda^4)
\]
which is always larger than the event horizon $\rho_h=1$. The remaining roots of (\ref{stella_nc}) cannot be candidates for the photon sphere since 
they must approach zero as $\lambda\to 0$ and therefore they have an asymptotic behaviour of the form $\rho_j(\lambda)\approx\lambda^{p_j}b_{0,j}$ 
with $p_j>0$ and $b_{0,j}\neq 0$ for all $j=1,2,3$. The relation between the impact parameter and the distance of closest approach is
\begin{equation}\label{2stella_nc}
\widetilde{b}=\rho_0\sqrt{\frac{\rho_0^2+\lambda}{\rho_0^2-(1+\lambda)\rho_0+\lambda}}. 
\end{equation}
Note that in the limit $\lambda\to 0$ the above relation reproduces correctly (\ref{ip_sch}) in the classic Schwarzschild case. Expanding 
(\ref{2stella_nc}) around $\rho_0=\rho_{3/2}$ we find $\widetilde{b}=\widetilde{b}_{cr}+\alpha_1(\rho_0-\rho_{3/2})^2
+\mathcal{O}(\rho_0-\rho_{3/2})^3$ with
\[
\widetilde{b}_{cr}=\frac{3\sqrt{3}}{2}+\frac{5\sqrt{3}}{6}\lambda+\cdots,\quad
\alpha_1=\frac{1}{2}\widetilde{b}^{''}(\rho_{3/2})=\left.\left\{-\frac{B^{'}}{2B\sqrt{B}}+\frac{\rho_0}{2\sqrt{B}}\left[\frac{3}{4}
\left(\frac{B^{'}}{B}\right)^2-\frac{B^{''}}{2B}\right]\right\}\right|_{\rho_0=\rho_{3/2}}=2\sqrt{3}-\frac{46\sqrt{3}}{27}\lambda+\cdots
\]
Note that as in the Schwarzschild case $\widetilde{b}^{'}(\rho_{3/2})=0$ because
$\widetilde{b}^{'}(\rho_0)=p(\rho_0)/(2\sqrt{\rho_0^2+\lambda}[\rho_0^2-(1+\lambda)\rho_0+\lambda]^{3/2})$ and $\rho_{3/2}$ is a root of the 
polynomial $p(\rho_0)$. The deflection angle due to a holographic screen in the weak field limit can be computed from (\ref{form_gen}) with 
$\rho,\rho_0 \gg 1$. Taking into account that $B(\rho)=1-(1+\lambda)/\rho+\cdots$ we find that
\[
\left(\frac{\rho}{\rho_0}\right)^2\frac{B(\rho_0)}{B(\rho)}-1=\left(\frac{\rho}{\rho_0}\right)^2\left[1-\frac{1+\lambda}{\rho_0}+\cdots\right]
\left[1+\frac{1+\lambda}{\rho}+\cdots\right]=\left[\left(\frac{\rho}{\rho_0}\right)^2-1\right]
\left[1-\frac{(1+\lambda)\rho}{\rho_0(\rho+\rho_0)}+\cdots\right]
\]
and
\[
\frac{\sqrt{A(\rho)}}{\rho\sqrt{\left(\frac{\rho}{\rho_0}\right)^2\frac{B(\rho_0)}{B(\rho)}-1}}=
\frac{1}{\rho\sqrt{\left(\frac{\rho}{\rho_0}\right)^2-1}}\left[1+\frac{1+\lambda}{2\rho}+\frac{(1+\lambda)\rho}{2\rho_0(\rho+\rho_0)}+\cdots\right]
\]
where $A=B^{-1}$. Let $h(\rho)=\rho^{-1}[(\rho/\rho_0)^2-1]^{-1/2}$. Since 
\[
\int_{\rho_0}^{\infty}h(\rho)d\rho=\frac{\pi}{2},\quad 
\int_{\rho_0}^{\infty}\frac{h(\rho)}{\rho}d\rho=\frac{1}{\rho_0},\quad
\int_{\rho_0}^{\infty}\frac{\rho h(\rho)}{\rho+\rho_0}d\rho=1,
\]
we find that in the weak field limit and at the first order in $1/\rho_0$ the deflection angle is related to the distance of closest approach through the following relation
\[
\Delta\varphi(\rho_0)=\frac{2(1+\lambda)}{\rho_0}+\cdots. 
\]
To treat strong gravitational lensing we will adopt the method developed by \cite{Bozza2}. First of all, we introduce new variables $y=B(\rho)$ and 
$z=(1-y)/(1-y_0)$ with $y_0=B(\rho_0)$. Then, $\rho$ can be expressed as a function of $z$ as follows
\[
\rho(z)=\frac{1+\lambda\pm\sqrt{(1+\lambda)^2-4\lambda(1-y_0)^2(1-z)^2}}{2(1-y_0)(1-z)}. 
\]
Since for $\lambda\to 0$ we should get as in the Schwarzschild case $\rho(z)=1/[(1-y_0)(1-z)]$, we must chose the positive sign in the above 
expression. The formula for the deflection angle can now be written as
\begin{equation}\label{2dot_nc}
\Delta\varphi(\rho_0)=-\pi+\int_0^1 R(z,\rho_0)f(z,\rho_0)~dz 
\end{equation}
with
\begin{equation}\label{RF}
R(z,\rho_0)=\frac{2(1-y_0)\rho_0}{\rho^2(z)B^{'}(\rho)},\quad 
f(z,\rho_0)=\left[y_0-[y_0+(1-y_0)z]\frac{\rho_0^2}{\rho^2(z)}\right]^{-1/2},
\end{equation}
where the prime denotes differentiation with respect to $\rho$, Note that the function $R$ does not exhibit singularities for any value of 
$z$ and $\rho_0$ while $f$ becomes singular as $z\to 0$ since $\rho(0)=\rho_0$. Expanding the argument of the square root in $f$ at the second 
order we find $f(z,\rho_0)\approx f_0(z,\rho_0)=[\alpha(\rho_0)z+\beta(\rho_0)z^2]^{-1/2}$ with
\begin{eqnarray*}
\alpha(\rho_0)&=&\frac{1-B(\rho_0)}{\rho_0 B^{'}(\rho_0)}[2B(\rho_0)-\rho_0 B^{'}(\rho_0)]=\frac{p(\rho_0)}{\rho_0(\rho_0^4-\lambda)},\\
\beta(\rho_0)&=&\frac{[1-B(\rho_0)]^2}{\rho_0^2 (B^{'}(\rho_0))^3}[2\rho_0(B^{'}(\rho_0))^2-3B(\rho_0)B^{'}(\rho_0)-\rho_0B(\rho_0)B^{''}(\rho_0)]. 
\end{eqnarray*}
A closer inspection of $\alpha$ shows that it becomes zero when $\rho=\rho_{3/2}$ and therefore the integral of $f$ will diverge logarithmically. 
Let us rewrite the integrand in (\ref{2dot_nc}) as $R(z,\rho_0)f(z,\rho_0)=R(0,\rho_{3/2})f_0(z,\rho_{3/2})+g(z,\rho_0)$ with 
$g(z,\rho_0)=R(z,\rho_0)f(z,\rho_0)-R(0,\rho_{3/2})f_0(z,\rho_{3/2})$ where
\[
R(0,\rho_{3/2})=\frac{2(\rho_{3/2}^2+\lambda)}{\rho_{3/2}^2-\lambda}=2+\frac{16}{9}\lambda+\cdots 
\]
tends correctly to the value $2$ for $\lambda\to 0$ as one would expect in the classic Schwarzschild case. Then, the integral for the deflection angle 
can be written as $\Delta\varphi(\rho_0)=-\pi+I_D(\rho_0)+I_R(\rho_0)$ where
\[
I_D(\rho_0)=R(0,\rho_{3/2})\int_0^1 f_0(z,\rho_{3/2})~dz,\quad
I_R(\rho_0)=\int_0^1 g(z,\rho_0)~dz 
\]
and the subscripts $D$ and $R$ stay for divergence and regular, respectively. The first integral admits the following exact solution
\[
\int_0^1 f_0(z,\rho_{3/2})~dz=\frac{2}{\sqrt{\beta}}\ln{\frac{\sqrt{\beta}+\sqrt{\alpha+\beta}}{\sqrt{\alpha}}}. 
\]
Expanding $\alpha$ and $\beta$ around $\rho_{3/2}$ we obtain
\begin{equation}\label{ID}
I_D(\rho_0)=-a\ln{\left(\frac{\rho_0}{\rho_{3/2}}-1\right)}+b_D+\mathcal{O}(\rho_0-\rho_{3/2}) 
\end{equation}
with
\[
a=\frac{R(0,\rho_{3/2})}{\sqrt{\beta(\rho_{3/2})}}=2+\frac{40}{27}\lambda+\cdots,\quad
b_D= \frac{R(0,\rho_{3/2})}{\sqrt{\beta(\rho_{3/2})}}\ln{\frac{2[1-B(\rho_{3/2})]}{\rho_{3/2}B^{'}(\rho_{3/2})}}
=2\ln{2}+\left(\frac{40}{27}\ln{2}+\frac{16}{9}\right)\lambda+\cdots
\]
The regular term in the deflection angle can be found by expanding the integral $I_R(\rho_0)$ in powers of $\rho_0-\rho_{3/2}$ as follows
\[
I_R(\rho_0)=\sum_{n=0}^{\infty}\frac{(\rho_0-\rho_{3/2})^n}{n!}\int_0^1\left.\frac{\partial^n g}{\partial \rho_0^n}\right|_{\rho_0=\rho_{3/2}}dz
=\int_0^1 g(z,\rho_{3/2})~dz+\mathcal{O}(\rho_0-\rho_{3/2}).
\]
Hence, the additional correction to be added to the term $-\pi+b_D$ is represented by $b_R=I_R(\rho_{3/2})$ and in the strong field limit the formula 
for the deflection angle reads
\[
\Delta\varphi(\rho_0)= -a\ln{\left(\frac{\rho_0}{\rho_{3/2}}-1\right)}+b_D+b_R\mathcal{O}(\rho_0-\rho_{3/2}).
\]
In this case it is not possible to give an analytical result for the integral representing $b_R$ but we can construct an expansion in the parameter 
$\lambda$. To this purpose note that
\begin{eqnarray*}
f_0(z,\rho_{3/2})&=&\frac{1}{z}-\frac{4}{27 z}\lambda+\cdots,\quad
R(z,\rho_{3/2})=2+\frac{8}{9}(3z^2-6z+2)\lambda+\cdots,\\
f(z,\rho_{3/2})&=&\frac{\sqrt{3}}{z\sqrt{3-2z}}-\frac{4\sqrt{3}(3z^2-6z+1)}{27z\sqrt{3-2z}}\lambda+\cdots. 
\end{eqnarray*}
Taking into account that $g(z,\rho_{3/2})=g_1(z)+g_2(z)\lambda+\cdots$ where
\[
g_1(z)=\frac{2\sqrt{3}-2\sqrt{3-2z}}{z\sqrt{3-2z}},\quad
g_2(z)=\frac{8(6\sqrt{3}z^2-12\sqrt{3}z+5\sqrt{3}-5\sqrt{3-2z})}{27z\sqrt{3-2z}},
\]
and integrating we obtain $b_R=b_{R,Sch}+\kappa\lambda+\cdots$ where 
\[
b_{R,Sch}=\ln{36}-\mbox{arctanh}{(\sqrt{3}/3)}=0.9496,\quad
\kappa=\frac{40}{27}\ln{6}+\frac{8(2\sqrt{3}-9)}{27}-\frac{80}{27}\mbox{arctanh}{(\sqrt{3}/3)}=-2.5770.
\]
Note that $b_{R,Sch}$ is in agreement with the numerical value found by \cite{Bozza2} for the classic Schwarzschild case.

\begin{figure}\label{Holo50}
\includegraphics[scale=0.5]{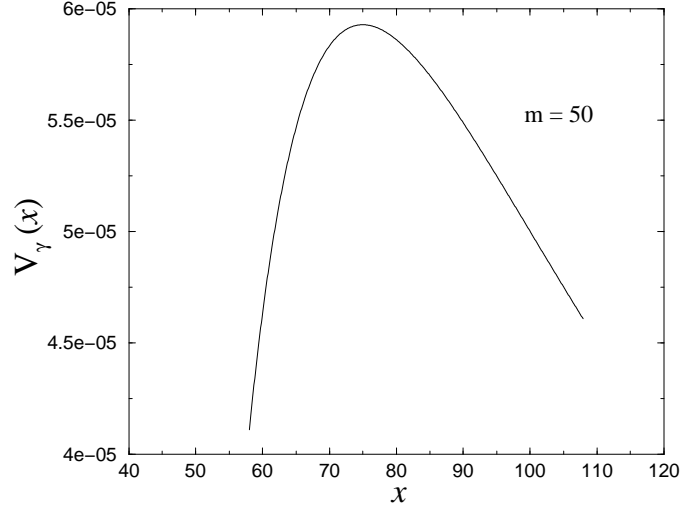}
\caption{\label{fig3.4}
The rescaled ``photon potential'' proportional to $\widetilde{V}$ for the Holographic screen metric
showing the photon unstable orbit (maximum) for a large $m$.
}
\end{figure}
 
\subsection{The extreme case}
Let $M=M_p$. Then, the Cauchy and event horizon coincide at $r_{\pm}=L_p$. Introducing the rescaling $\rho=r/L_p$ the metric function $B$ is now 
given by $B(\rho)=(\rho-1)^2/(\rho^2+1)$. The radius of the photon sphere is obtained by solving the equation $B^{'}(\rho)/B(\rho)=2/\rho$ which 
gives rise to the cubic equation $q(\rho)=\rho^3-2\rho^2-1=0$. This equation has two immaginary roots and one real root located at 
\[
\rho_f=\frac{2}{3}+\frac{(172+12\sqrt{177})^{2/3}+16}{(172+12\sqrt{177})^{1/3}}=2.2057.
\]
The impact parameter and the distance of closest approach are related to each other through $\widetilde{b}=\rho_0\sqrt{\rho_0^2+1}/(\rho_0-1)$. 
An expansion around the point $\rho_0=\rho_f$ gives $\widetilde{b}=\widetilde{b}_{cr}+\widetilde{\alpha}_1(\rho_0-\rho_f)^2
+\mathcal{O}(\rho_0-\rho_f)^2$ with 
\[
\widetilde{b}_{cr}=\frac{\rho_f\sqrt{\rho_f^2+1}}{\rho_f-1}=4.4304,\quad
\widetilde{\alpha}_1=\frac{3\rho_f^3+3\rho_f+2}{2(\rho_f-1)^3(\rho_f+1)^{3/2}}=0.8199 
\]
To obtain the weak deflection limit for the angle $\Delta\varphi$ we suppose that $\rho,\rho_0 \gg 1$ and proceding exactly as we did in the 
nonextreme  case we find at the first order in $1/\rho_0$
\[
\Delta\varphi(\rho_0)=\frac{4}{\rho_0}+\cdots 
\]
To study the strong gravitational lensing we change variables according to $y=B(\rho)$ and $z=(1-y)/(1-y_0)$. Solving the first equation for $\rho$ 
we obtain $\rho(y)=(1\pm\sqrt{2y-y^2})/(1-y)$. Since $\rho\to 1$ as $y\to 0$ and $\rho\to\infty$ for $y\to 1$, we have to pick the solution with 
the plus sign which can be expressed in terms of the variable $z$ as 
\[
\rho(z)=\frac{1+\sqrt{1-(1-y_0)^2(1-z)^2}}{(1-y_0)(1-z)}. 
\]
At this point the deflection angle will be given by (\ref{2dot_nc}) with the functions $R$ and $f$ formally given by (\ref{RF}). In the present case 
the coefficients $\alpha$ and $\beta$ entering in the expansion of $f$ are given by
\[
\alpha(\rho_0)=\frac{2q(\rho_0)}{(\rho_0+1)(\rho_0^2+1)},\quad 
\beta(\rho_0)=\frac{\rho_0^4-4\rho_0^3-2\rho_0^2-4\rho_0-3}{(1-\rho_0)(\rho_0+1)^3}. 
\]
Note that $\alpha$ vanishes whenever the distance of closest approach coincides with the radius of the photon sphere. As in the nonextreme case we 
rewrite the deflection angle as $\Delta\varphi(\rho_0)=-\pi+I_D(\rho_0)+I_R(\rho_0)$. The integral $I_D$ can be expanded about $\rho_f$ and one 
obtains formally an expression as (\ref{ID}) where $\rho_{3/2}$ is replaced by $\rho_f$ and 
\[
a=\frac{2(\rho_f^2+1)}{\rho_f-1}\sqrt{\frac{1-\rho_f^2}{\rho_f^4-4\rho_f^3-2\rho_f^2-4\rho_f-3}}=2.9941,\quad
b_D=a\ln{\frac{2(\rho_f^2+1)}{\rho_f-1}}=6.8120 
\]
To compute the coefficient $b_R$ we need the following quantities
\[
R(z,\rho_0)=\frac{4\rho_0^2(\rho_0^2+1)}{\rho_0^4+(\rho_0^2+1)\sqrt{(\rho_0^2-1)^2+4\rho_0^2 z(2-z)}+2\rho_0^2(4z-2z^2-1)+1},\quad
R(0,\rho_0)=\frac{2(\rho_0^2+1)}{\rho_0^2-1}. 
\]
However, the integral giving $b_R$ can be solved only numerically and we find $b_R=-1.0217$

\begin{figure}\label{Holocrit}
\includegraphics[scale=0.5]{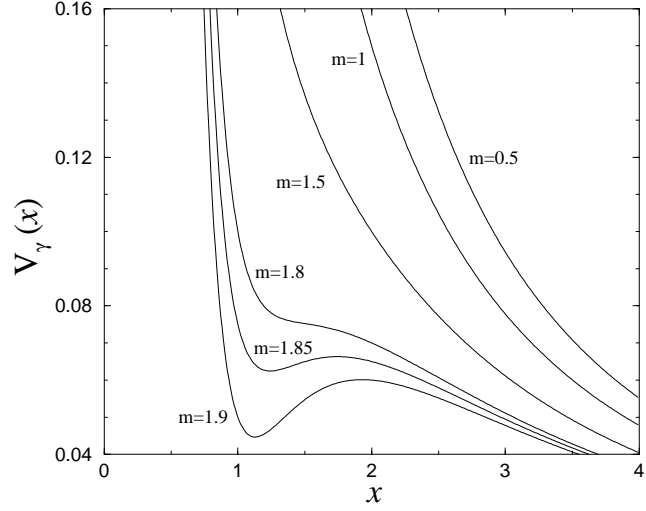}
\caption{\label{fig3.5}
The rescaled ``photon potential'' proportional to $\widetilde{V}$ for the Holographic screen metric
displaying the naked singularity case where a narrow range of the parameter $m$ leads to a
local minimum allowing bound states of photons.
}
\end{figure}

\section{Gravitational lensing for noncommutative geometry inspired wormholes}
Noncommutative geometry inspired wormholes are solutions of the Einstein field equations obtained by assuming that the 
mass/energy distribution is a Gaussian of the form
\[
\rho(r)=\frac{M}{(4\pi\theta)^{3/2}}e^{-r^2/(4\theta)}, 
\]
where $\sqrt{\theta}$ is the matter distribution width and it defines the scale where the spacetime coordinates should be replaced by some 
noncommuting coordinate operators in a suitable Hilbert space \cite{PN1}. Here, $M$ is the total mass and is given by the integral 
$M=4\pi\int_0^\infty r^2\rho(r)~dr$. The gravitational source is then modelled by a fluid-type energy-momentum tensor of the form 
$T^\mu{}_\nu=\mbox{diag}(\rho(r),-p_r(r),-p_\bot(r),-p_\bot(r))$ with $p_r$ and $p_\bot$ the radial and tangential pressures, respectively, 
together with the condition $T^\mu{}_{\nu;\mu}=0$. If we look for a metric such that it is spherically symmetric, static, and asymptotically flat, 
then we can write the corresponding line element as
\[
ds^2=e^{2\Phi(r)}dt^2-\left(1-\frac{2m(r)}{r}\right)^{-1}dr^2-r^2 d\Omega^2, 
\]
where $\Phi(r)$ and $m(r)$ are the so-called red-shift and shape functions, which must be determined by solving Einstein equations. If we assume 
that $p_r(r)=-m(r)/(4\pi r^3)$ and $m(r)=\int_0^r u^2\rho(u)~du$, then we obtain the following line element
\begin{equation}\label{sw}
ds^2=dt^2-\left(1-\frac{4M}{\sqrt{\pi}r}\gamma\left(\frac{3}{2};\frac{r^2}{4\theta}\right)\right)^{-1}dr^2-r^2 d\Omega^2,\quad
\gamma\left(\frac{3}{2};\frac{r^2}{4\theta}\right)=\int_0^{r^2/(4\theta)}\sqrt{s}e^{-s}~ds 
\end{equation}
describing a wormhole. Properties of the metric (\ref{sw}) have been investigated in \cite{GAR}. Here, we extend the results of \cite{GAR} by 
studying gravitational lensing in the presence of the noncommutative geometry inspired wormhole given by (\ref{sw}). A throat will exist if 
$2m(r_t)=r_t$ for some $r_t>\sqrt{\theta}$. As in \cite{Piero1} we will have two distinct throats $r_{t,+}>r_{t,-}$ if $M>M_0=1.9042\sqrt{\theta}$, 
two coinciding throats $r_{t,+}=r_{t,-}=r_e=3.0226\sqrt{\theta}$ (extreme case) whenever $M=M_0$ and no throats for $M<M_0$. We will consider only 
weak gravitational lensing for the nonextreme and extreme cases, since in both regimes there is no photon sphere and therefore a light ray 
approaching the wormhole will be either deflected or disappear into the throat. The absence of a photon sphere is due to the fact that the equation 
$2/x=B^{'}(x)/B(x)$ can never be satisfied because the metric coefficient $B$ is a constant function.
\subsection{Nonextreme case}
\begin{figure}\label{darkw}
\includegraphics[scale=0.5]{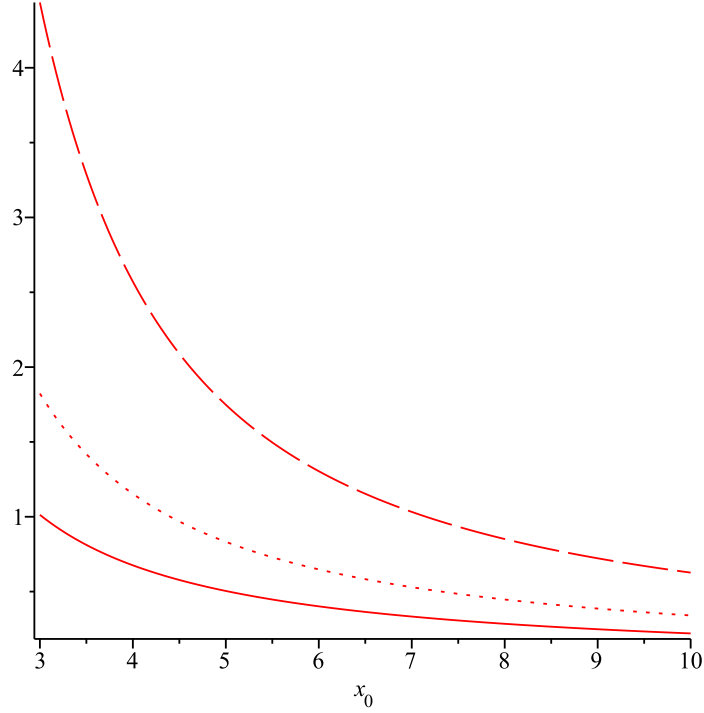}
\caption{\label{darkmw}
Plot of the weak field approximation (\ref{defang}) as a function of the closest distance of approach for $\alpha=2.2$ (solid), 
$\alpha=3.2$ (dot), $\alpha=5.2$ (longdash). 
}
\end{figure}

Let us rescale the time and spatial coordinates and the mass according to $\widetilde{t}=t/(2\sqrt{\theta})$, $x=r/(2\sqrt{\theta})$, and 
$\alpha=r_S/\sqrt{\pi\theta}$ where $r_S$ denotes the classic Schwarzschild radius. Then, the nonextremality condition reads $\alpha>\alpha_0=2.1$ 
and the metric (\ref{sw}) can be brought into the form $ds^2=4\theta d\widetilde{s}^2$ with 
\[
d\widetilde{s}^2=d\widetilde{t}^2-A(x)dx^2-x^2 d^2\Omega,\quad 
A(x)=\left[1-\frac{\alpha}{x}\gamma\left(\frac{3}{2};x^2\right)\right]^{-1}.
\]
Let $x_0$ denote the distance of closest approach. Further suppose that $x,x_0\gg 1$. Since the metric coefficient $B(x)=1$, the integral expressing 
the deflection angle as a function of $x_0$ simplifies to
\[
\Delta\varphi(x_0)=-\pi+2\int_{x_0}^\infty \frac{\sqrt{A(x)}~dx}{x\sqrt{(x/x_0)^2-1}}.
\]
Taking into account that $\gamma(3/2;x^2)=\sqrt{\pi}/2-\Gamma(3/2;x^2)$ where $\Gamma(\cdot;\cdot)$ denotes the upper incomplete Gamma function 
and using $6.5.32$ in \cite{abra} we get the following asymptotic expansion for the lower incomplete Gamma function
\[
\gamma\left(\frac{3}{2};x^2\right)=\frac{\sqrt{\pi}}{2}-xe^{-x^2}\left[1+\mathcal{O}(x^{-2})\right], 
\]
which in turn allows to construct an asymptotic expansion for $\sqrt{A(x)}$ represented by
\begin{equation}\label{hertz}
\sqrt{A(x)}=\left[1-\frac{\alpha\sqrt{\pi}}{2x}+\alpha e^{-x^2}+\mathcal{O}\left(\frac{e^{-x^2}}{x^2}\right)\right]^{-1/2}. 
\end{equation}
To further expand the above expression we make the substitution $x^2=\ln{u}$ and we obtain
\[
\sqrt{A(u)}=\left[1-\frac{\alpha\sqrt{\pi}}{2\sqrt{\ln{u}}}+\frac{\alpha}{u}+\mathcal{O}\left(\frac{1}{u\ln{u}}\right)\right]^{-1/2}
=f_0(u)+\frac{f_1(u)}{u}+\frac{f_2(u)}{u^2}+\mathcal{O}\left(\frac{1}{u^3}\right)
\]
with
\[
f_0(u)=\sqrt{\frac{2\sqrt{\ln{u}}}{2\sqrt{\ln{u}}-\alpha\sqrt{\pi}}},\quad 
f_1(u)=-\frac{\alpha f_0(u)\sqrt{\ln{u}}}{2\sqrt{\ln{u}}-\alpha\sqrt{\pi}},\quad
f_2(u)=\frac{3\alpha^2 f_0(u)\ln{u}}{2(2\sqrt{\ln{u}}-\alpha\sqrt{\pi})^2}.
\]
A further expansion of the functions $f_0,f_1,f_2$ gives
\[
f_0(u)=1+\frac{\alpha\sqrt{\pi}}{4\sqrt{\ln{u}}}+\frac{3\alpha^2\pi}{32\ln{u}}+\frac{5\alpha^3\pi\sqrt{\pi}}{128\ln{u}\sqrt{\ln{u}}}+
\mathcal{O}\left(\frac{1}{u\ln{u}}\right),\quad
\frac{f_1(u)}{u}=-\frac{\alpha}{2u}-\frac{3\alpha^2\sqrt{\pi}}{8u\sqrt{\ln{u}}}+ \mathcal{O}\left(\frac{1}{u\ln{u}}\right)
\]
whereas $f_2/u^2$ is of order $\mathcal{O}(1/(u\ln{u}))$. Finally, going back to the variable $x$ yields the following asymptotic expansion 
\[
\sqrt{A(x)}=1+\frac{\alpha\sqrt{\pi}}{4x}+\frac{3\alpha^2\pi}{32 x^2}+\frac{5\alpha^3\pi\sqrt{\pi}}{128 x^3}-\frac{\alpha}{2}e^{-x^2}
-\frac{3\alpha^2\sqrt{\pi}}{8}\frac{e^{-x^2}}{x}+\mathcal{O}\left(\frac{e^{-x^2}}{x^2}\right).
\]
Let $h(x)=2x^{-1}[(x/x_0)^2-1]^{-1/2}$ so that the integral giving the deflection angle can be written in the more compact form
$\Delta\varphi(x_0)=-\pi+\int_{x_0}^{\infty}h(x)\sqrt{A(x)}~dx$. Then, we get
\[
\int_{x_0}^{\infty}h(x)~dx=\pi,\quad \int_{x_0}^{\infty}\frac{h(x)}{x}~dx=\frac{2}{x_0},\quad
\int_{x_0}^{\infty}\frac{h(x)}{x^2}~dx=\frac{\pi}{2x_0^2},\quad
\int_{x_0}^{\infty}\frac{h(x)}{x^3}~dx=\frac{4}{3x_0^3}.
\]
Moreover, $\int_{x_0}^{\infty}h(x)e^{-x^2}~dx=\pi[1-\mbox{erf}(x_0)]$ where $\mbox{erf}(\cdot)$ is the error function for which we can construct the 
asymptotic expansion
\[
\mbox{erf}(x_0)=1-\frac{e^{-x_0^2}}{\sqrt{\pi}x_0}+ \mathcal{O}\left(\frac{e^{-x_0^2}}{x_0^2}\right),
\]
by means of relations $7.1.2$ and $7.1.23$ in \cite{abra} and
\[
\int_{x_0}^{\infty}h(x)\frac{e^{-x^2}}{x}~dx=x_0 e^{-x_0^2/2}\left[K_1\left(\frac{x_0^2}{2}\right)-K_1\left(\frac{x_0^2}{2}\right)\right]=
\mathcal{O}\left(\frac{e^{-x_0^2}}{x_0^2}\right) 
\]
where we used the asymptotic expansion $9.7.2$ for the modified Bessel functions given in \cite{abra}. Putting things together we find that
\begin{equation}\label{defang}
\Delta\varphi(x_0)=\frac{\alpha\sqrt{\pi}}{2x_0}\left(1-e^{-x_0^2}\right)+\frac{3\alpha^2\pi^2}{64 x_0^2}+\frac{5\alpha^3\pi\sqrt{\pi}}{96 x_0^3}+
\mathcal{O}\left(\frac{e^{-x_0^2}}{x_0^2}\right). 
\end{equation}
We plotted the behaviour of (\ref{defang}) in Figure~\ref{darkmw}. Going back to the unscaled distance of closest approach the deflection angle in the weak field limit reads
\[
\Delta\varphi(r_0)=\frac{2M}{r_0}\left(1-e^{-r_0^2/(4\theta)}\right)+\frac{3\pi M^2}{4 r_0^2}+\frac{10 M^3}{3 r_0^3}+
\mathcal{O}\left(\frac{e^{-r_0^2/(4\theta)}}{r_0^2/(4\theta)}\right). 
\]
\subsection{Extreme case}
In this case the throats coincide at $x_e$ and we can analyze the metric coefficient $A(x)$ as in \cite{wir_2010}. The line element (\ref{sw}) 
becomes after the usual rescaling $\widetilde{t}=t/(2\sqrt{\theta})$, $x=r/(2\sqrt{\theta})$, and $\alpha_0=2M_0/(\sqrt{\pi\theta})$
\[
ds^2_E=d\widetilde{t}^2-\frac{dx^2}{(x-x_e)^2 \phi(x)}-x^2 d\Omega^2, 
\]
where $\phi$ is a differentiable and not vanishing function in the interval $[0,+\infty)$. Moreover,
\[
\phi(x_e)=\frac{1}{2}f^{''}(x_e),\quad \phi^{'}(x_e)=\frac{1}{6}f^{'''}(x_e),\quad f(x)=1-\frac{\alpha}{x}\gamma\left(\frac{3}{2};x^2\right). 
\]
Using the software Maple we find the following numerical values $f^{''}(x_1)=0.5620$ and $f^{'''}(x_1)=-0.3732$. Concerning the weak field limit we 
can again use formula (\ref{defang}) with $\alpha$ replaced by $\alpha_0$ since asymptotically at infinity $1/\sqrt{f(x)}$ has the same asymptotic 
behaviour $\sqrt{A(x)}$.

\section{Gravitational lensing for noncommutative geometry inspired dirty black holes}
Dirty black holes are solutions of Einstein field equations in the presence of various classical matter fields such as electromagnetic fields, dilaton fields, axion fields, Abelian Higgs fields, non-Abelian gauge fields, etc. . We will study gravitational lensing 
for a dirty black hole inspired by noncommutative geometry and described by the line element (\ref{25.3}) with \cite{PN1}
\[
A(r)=\left[1-\frac{4M}{\sqrt{\pi}r}\gamma\left(\frac{3}{2};\frac{r^2}{4\theta}\right)\right]^{-1},\quad
B(r)=\frac{1}{A(r)}e^{-\frac{M}{\sqrt{\theta}}\left[1-\frac{2}{\sqrt{\pi}}\gamma\left(\frac{3}{2};\frac{r^2}{4\theta}\right)\right]} 
\]
with $M$ and $\sqrt{\theta}$ defined as in the previous section. Note that this line element represents a generalization of the noncommutative geometry inspired Schwarzschild metric derived in \cite{Piero1}.  After the usual rescaling $\widetilde{t}=t/(2\sqrt{\theta})$, 
$x=r/(2\sqrt{\theta})$, and $\alpha=r_S/\sqrt{\pi\theta}$ the metric functions $A$ and $B$ read
\[
A(x)=\left[1-\frac{\alpha}{x}\gamma\left(\frac{3}{2};x^2\right)\right]^{-1},\quad
B(x)=\frac{1}{A(x)}e^{-\frac{\alpha\sqrt{\pi}}{2}\left[1-\frac{2}{\sqrt{\pi}}\gamma\left(\frac{3}{2};x^2\right)\right]}.  
\]
The analysis of the existence of horizons $x_h$ can be performed by studying the roots of the equation $A(x)=0$. Unfortunately their positions 
can only be given implicitely as 
\begin{equation}\label{nc_1_star}
x_h=\alpha \gamma\left(\frac{3}{2};x^2_h\right).
\end{equation}
In particular, we will have the following scenarios
\begin{itemize}
\item 
two distinct horizons $x_{+}>x_{-}$ for $\alpha>\alpha_0=2.1486$ (non-extremal dirty black hole);
\item
one degenerate horizon at $x_e=x_{-}=x_{+}=1.5113$ for $\alpha=\alpha_0$ (extremal dirty black hole);
\item
no horizons for $0<\alpha<\alpha_0$ (dirty minigravastar).
\end{itemize}
Note that using (\ref{nc_1_star}) and $6.5.3$ in \cite{abra} we can express the event horizon in terms of the incomplete upper Gamma function as
\[
x_{+}=\frac{\alpha\sqrt{\pi}}{2}-\alpha\Gamma\left(\frac{3}{2};x^2_{+}\right). 
\]
Since $\Gamma(3/2;x^2)\to 0$ as $x\to\infty$, it can be easily checked that $x_{+}$ tends to $\alpha\sqrt{\pi}/2$ which corresponds correctly to 
$r_{+}\to 2M$. The radius of the photon sphere can be obtained by finding the roots of the equation
\begin{equation}\label{pse}
\frac{B^{'}(x)}{B(x)}=\frac{2}{x}. 
\end{equation}
Using $6.5.25$ in \cite{abra} yields $d\gamma(3/2;x^2)/dx=2x^2 e^{-x^2}$ and the radius $x_p$ of the photon sphere can be given implicitely by the 
following formula
\begin{equation}\label{phot_sph}
x_p=\frac{3\alpha\sqrt{\pi}}{4}-\left[\alpha x_p^3 e^{-x_p^2}+\frac{3}{2}\alpha\Gamma\left(\frac{3}{2};x^2_p\right)
-\alpha x^4_p e^{-x^2_p}+\alpha^2 x^3_p e^{-x^2_p}\gamma\left(\frac{3}{2};x^2_p\right)\right], 
\end{equation}
where we also applied $6.5.3$ in \cite{abra}. In the Schwarzschild limit $x\to\infty$ the above expression correctly reproduces the radius of the 
photon sphere at $3M$. Moreover, (\ref{phot_sph}) generalizes formula $(2.10)$ obtained by \cite{Ding} for the photon sphere of a noncommutative 
geometry inspired Schwarzschild black hole. 
\begin{figure}\label{dirtyw}
\includegraphics[scale=0.5]{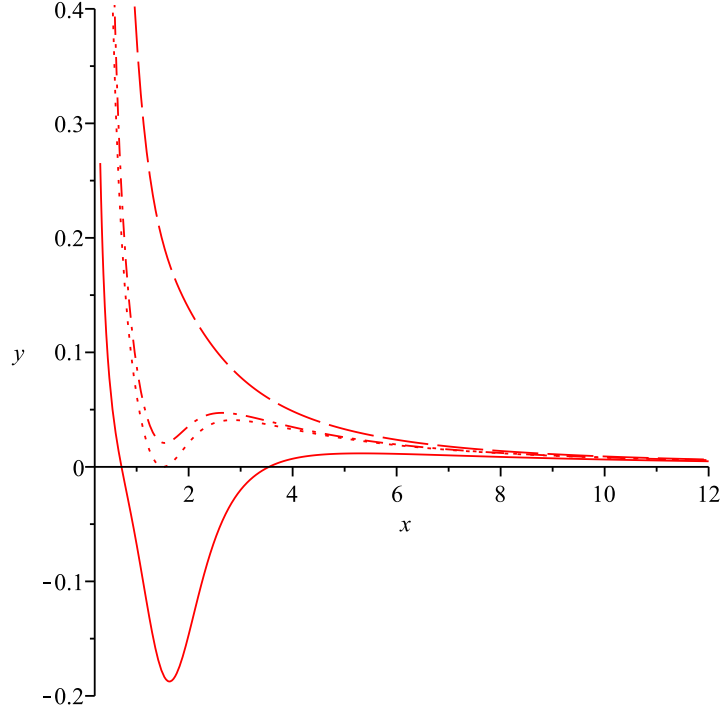}
\caption{\label{dirtymw}
Plot of (\ref{vtilde}) for the non-extremal case (solid line, $\alpha=4$), extremal case (dotted line, $\alpha=\alpha_0$), and the dirty 
minigravastar (dash dotted and long dashed lines for $\alpha=2$ and $\alpha=1$, respectively). It is interesting to observe that in the case of a 
minigravastar light will form bound states only when the parameter $\alpha$ varies in a certain interval. 
}
\end{figure}
Let $x_0$ denote the distance of closest approach. Then, the rescaled impact parameter is given by
\[
\widetilde{b}(x_0)=\frac{x_0}{\sqrt{B(x_0)}}=x_0  
e^{\frac{\alpha\sqrt{\pi}}{4}\left[1-\frac{2}{\sqrt{\pi}}\gamma\left(\frac{3}{2};x^2_0\right)\right]}
\left[1-\frac{\alpha}{x_0}\gamma\left(\frac{3}{2};x^2_0\right)\right]^{-1/2}.
\]
Since $\widetilde{b}^{'}(x_0)=[2B(x_0)-x_0 B^{'}(x_0)]/(2B^{3/2}(x_0))$, it will clearly vanish at $x_0=x_p$ and the impact parameter can be 
expanded in a neighbourhood of $x_p$ as 
\[
\widetilde{b}(x_0)= \widetilde{b}(x_p)+\frac{1}{2}\widetilde{b}^{''}(x_p)(x_0-x_p)^2+\mathcal{O}(x_0-x_p)^3,\quad
\widetilde{b}^{''}(x_p)=\frac{B^{'}(x_p)-x_p B^{''}(x_p)}{2B^{3/2}(x_p)}.
\]
Using $6.5.32$ in \cite{abra} it can be verified that $\widetilde{b}(x_p)\to x_p$ as $x_p\to\infty$, from which we recover correctly 
$b(r_0)=3\sqrt{3} M$ as one would expect in the case of the Schwarzschild metric. In what follows we will treat at the same time both the 
nonextreme and the extreme cases by letting $\alpha\geq \alpha_0$. Concerning the weak field limit of the deflection angle we will suppose that 
$x,x_0\gg 1$ in
\[
\Delta\varphi(x_0)=-\pi+2\int_{x_0}^\infty\frac{\sqrt{A(x)~dx}}{x\sqrt{\left(\frac{x}{x_0}\right)^2\frac{B(x_0)}{B(x)}-1}}. 
\]
An asymptotic expansion for $\sqrt{A(x)}$ has been already derived in the previous section and it is represented by (\ref{hertz}). Taking into account 
that
\[
B(x_0)=1-\frac{\alpha\sqrt{\pi}}{2x_0}+\alpha e^{-x_0^2}+\mathcal{O}\left(\frac{e^{-x_0^2}}{x_0^2}\right),\quad
\frac{1}{B(x)}=1+\frac{\alpha\sqrt{\pi}}{2x}+\mathcal{O}\left(\frac{1}{x^2}\right),
\]
we find that
\[
\frac{B(x_0)}{B(x)}=1+\frac{\alpha\sqrt{\pi}}{2}\left(\frac{1}{x}-\frac{1}{x_0}\right)+\alpha e^{-x_0^2}+\cdots 
\]
and hence
\[
\left(\frac{x}{x_0}\right)^2\frac{B(x_0)}{B(x)}-1=\left(\frac{x}{x_0}\right)^2-1-\frac{\alpha\sqrt{\pi}}{2}\frac{x(x-x_0)}{x_0^3}
+\alpha\frac{x^2}{x^2_0}e^{-x_0^2}+\cdots=\left[\left(\frac{x}{x_0}\right)^2-1\right]\Psi(x,x_0),
\]
where
\[
\Psi(x,x_0)=1-\frac{\alpha\sqrt{\pi}}{2}\frac{x}{x_0(x+x_0)}+\frac{\alpha x^2}{x^2-x_0^2}e^{-x_0^2}+\cdots.
\]
Finally, by means of (\ref{hertz}) we obtain the asymptotic expansion
\[
\frac{\sqrt{A(x)~dx}}{x\sqrt{\left(\frac{x}{x_0}\right)^2\frac{B(x_0)}{B(x)}-1}}=
\frac{1}{x\sqrt{\left(\frac{x}{x_0}\right)^2-1}}\left[1+\frac{\alpha\sqrt{\pi}}{4}\frac{x^2+x_0(x+x_0)}{x_0 x(x+x_0)}
-\frac{\alpha}{2}e^{-x^2}+\cdots\right].
\]
Let $h(x)=2x^{-1}[(x/x_0)^2-1]^{-1/2}$. Then, 
\[
\int_{x_0}^\infty h(x)~dx=\pi,\quad  \int_{x_0}^\infty h(x)\frac{x^2+x_0(x+x_0)}{x_0 x(x+x_0)}~dx=\frac{4}{x_0},\quad
\int_{x_0}^\infty h(x)e^{-x^2}~dx=\pi[1-\mbox{erf}(x_0)]=\sqrt{\pi}\frac{e^{-x_0^2}}{x_0}+\mathcal{O}\left(\frac{e^{-x_0^2}}{x_0^2}\right)
\]
and the deflection angle can be finally written as
\[
\Delta\varphi(x_0)=\frac{\alpha\sqrt{\pi}}{x_0}\left(1-\frac{e^{-x_0^2}}{2}\right)+\cdots 
\]
Rewriting the above result by means of the distance of closest approach $r_0$ as
\begin{equation}\label{wfldbh}
\Delta\varphi(r_0)=\frac{4M}{r_0}\left(1-\frac{e^{-\frac{r^2}{4\theta}}}{2}\right)+\cdots 
\end{equation}
we see that in the limit $r_0/\sqrt{2\theta}\to\infty$ it correctly reproduces the result one would expect for the classic Schwarzschild metric. 
From the above formula we see that the effect of noncommutative geometry is that of reducing the deflection angle. Last but not least, formula 
(\ref{wfldbh}) will also apply to the noncommutative geometry inspired Schwarzschild black hole since for large values of $x$ the metric of a 
noncommutative dirty black hole goes over into the metric of the aforementioned noncommutative black hole. Concerning the strong gravitational 
limit of the deflection angle we cannot introduce the transformations $y=B(x)$ and $z=(1-y)/(1-y_0)$ as in \cite{Bozza2} since the metric 
coefficient $B$ contains the lower incomplete gamma function in such a way that it results impossible to solve analytically the equation $y=B(x)$ 
for $x$. For this reason we will adopt the same choice as in \cite{Ding} and introduce a new variable $z=1-x_0/x$ in terms of which the integral 
giving the deflection angle can be expressed as $\Delta\varphi(x_0)=-\pi+I(x_0)$ with
\[
I(x_0)=\int_0^1 R(z,x_0)f(z,x_0)~dz,\quad
R(z,x_0)=2\sqrt{A(z,x_0)B(z,x_0)},\quad f(z,x_0)=\left[B(x_0)-(1-z)^2 B(z,x_0)\right]^{-1/2},\quad
\]
and
\[
B(z,x_0)=e^{-\frac{\alpha\sqrt{\pi}}{2}\left[1-\frac{2}{\sqrt{\pi}}\gamma\left(\frac{3}{2};\frac{x_0^2}{(1-z)^2}\right)\right]}
\left[1-\frac{\alpha(1-z)}{x_0}\gamma\left(\frac{3}{2};\frac{x_0^2}{(1-z)^2}\right)\right].
\]
At this point some comments are in order. First of all, for the noncommutative geometry inspired Schwarzschild black hole \cite{Piero1} we would 
have $R(z,x_0)=2$. Moreover, $R$ is a regular function of $z$ and $x_0$ with
\[
R(0,x_0)=2 e^{-\frac{\alpha\sqrt{\pi}}{4}\left[1-\frac{2}{\sqrt{\pi}}\gamma\left(\frac{3}{2};x_0^2\right)\right]},\quad
R(1,x_0)=2,
\]
where for $R(1,x_0)$ we used the result $\lim_{z\to 1}\gamma(3/2;x_0^2/(1-z)^2)=\lim_{x\to\infty}\gamma(3/2;x^2)=\sqrt{\pi}/2$. The result 
$R(1,x_0)=2$ is not surprising since for $x\to\infty$ the metric under consideration goes over into the classic Schwarzschild solution. A closer 
inspection of the function $f$ reveals that there is a singularity at $z=0$. Expanding the argument of the square root in $f$ to the second order 
in $z$ we get $f(z,x_0)\approx f_0(z,x_0)=[\widetilde{\alpha}(x_0)z^2+\widetilde{\beta}(x_0)z]^{-1/2}$ where
\[
\widetilde{\alpha}(x_0)=2B(x_0)-x_0 B^{'}(x_0),\quad
\widetilde{\beta}(x_0)=-B(x_0)+x_0 B^{'}(x_0)-\frac{x_0^2}{2}B^{''}(x_0),\quad ^{'}=\frac{d}{dx}.
\]
We immediately see that at the photon sphere $\widetilde{\alpha}(x_p)=0$ and therefore, $f$ diverges as $z^{-1}$ there, whereas for $x_0>x_p$ the 
function $f$ will behave as $z^{-1/2}$ which is clearly integrable at $z=0$. For $x_0<x_p$ every photon will be captured by the dirty black hole. As in 
\cite{Bozza2} we split the integral for the deflection angle into a regular and divergent part, respectively, that is $I(x_0)=I_D(x_0)+I_R(x_0)$ where 
$I_D(x_0)=R(0,x_p)\int_0^1 f_0(z,x_0)~dz$ contains the divergence and $I_R(x_0)=\int_0^1 g(z,x_0)~dz$ with 
$g(z,x_0)=R(z,x_0)f(z,x_0)-R(0,x_p)f_0(z,x_0)$ is regular since we subtracted the divergence. The integral $I_D$ can be computed analytically to give 
\[
I_D(x_0)=\frac{2R(0,x_p)}{\sqrt{\widetilde{\beta}(x_0)}}\log{\frac{\sqrt{\widetilde{\beta}(x_0)}+\sqrt{\widetilde{\alpha}(x_0)
+\widetilde{\beta}(x_0)}}{\sqrt{\widetilde{\alpha}(x_0)}}}.
\]
Expanding $\widetilde{\alpha}$ and $\widetilde{\beta}$ around the radius $x_p$ of the photon sphere we find
\[
\widetilde{\alpha}(x_0)=\widetilde{\alpha}_1(x_p)(x_0-x_p)+\mathcal{O}(x_0-x_p)^2,\quad
\widetilde{\beta}(x_0)=\widetilde{\beta}_0(x_p)+\widetilde{\beta}_1(x_p)(x_0-x_p)+\mathcal{O}(x_0-x_p)^2,
\]
where
\[
\widetilde{\alpha}_1(x_p)=B^{'}(x_p)-x_p B^{''}(x_p),\quad
\widetilde{\beta}_0(x_p)=-B(x_p)+x_p B^{'}(x_p)-\frac{x_p^2}{2}B^{''}(x_p),\quad
\widetilde{\beta}_1(x_p)=-\frac{x_p^2}{2}B^{'''}(x_p).
\]
Hence, the divergent part $I_D$ of the integral can be expanded according to $I_D=-a\log{(x_0-x_p)}+b_D+\mathcal{O}(x_0-x_p)$ where
\[
a=\frac{R(0,x_p)}{\sqrt{\widetilde{\beta}_0(x_p)}},\quad b_D=a\log{\frac{4\widetilde{\beta}_0(x_p)}{\widetilde{\alpha}_1(x_p)}}. 
\]
Following \cite{Bozza2} the deflection angle in the strong field limit will be given by $\Delta\varphi(x_0)=-a\log{(x_0-x_p)}+b+\mathcal{O}(x_0-x_p)$ 
with $b=-\pi+b_D+b_R$ where $b_R=I_R(x_p)=\int_0^1 g(z,x_p)~dz+\mathcal{O}(x_0-x_p)$. Unfortunately, $b_R$ can be only computed numerically. In 
Table~\ref{table2} we present some typical numerical values for the parameters $a$, $b_R$, and $b_D$ for the extreme case ($\alpha=\alpha_0$) and 
the non extreme case ($\alpha>\alpha_0$). For the treatment of the strong field limit in the presence of the noncommutative geometry inspired 
Schwarzschild meric we refer to \cite{Ding}
\begin{table}[ht]
\caption{Numerical values of the horizons $x_{\pm}$, the photon sphere $x_p$, and the coefficients $a$, $b_D$, $b_R$ for $\alpha\geq\alpha_0$. The 
first line corresponds to the extreme case.}  
\begin{center}
\begin{tabular}{ | l | l | l | l|l|l|l|}
\hline
$\alpha$ & $x_{-}$  & $x_{+}$ & $x_p$  & $a$    & $b_D$  & $b_R$  \\ \hline
2.1486   & 1.5113   & 1.5113  & 2.8526 & 2.0339 & 3.5420 & 0.3675 \\ \hline
2.1800   & 1.3583   & 1.6813  & 2.8952 & 2.0270 & 3.5600 & 0.3721 \\ \hline
2.2000   & 1.3186   & 1.7321  & 2.9222 & 2.0233 & 3.5724 & 0.3742 \\ \hline
2.3000   & 1.1972   & 1.9097  & 3.0566 & 2.0108 & 3.6406 & 0.3785 \\ \hline
2.4000   & 1.1208   & 2.0433  & 3.1901 & 2.0048 & 3.7153 & 0.3787 \\ \hline
2.5000   & 1.0630   & 2.1596  & 3.3233 & 2.0017 & 3.7914 & 0.3754 \\ \hline
2.6000   & 1.0161   & 2.2664  & 3.4563 & 2.0007 & 3.8682 & 0.3711 \\ \hline
2.7000   & 0.9765   & 2.3673  & 3.5892 & 2.0001 & 3.9424 & 0.3622 \\ \hline
2.8000   & 0.9421   & 2.4643  & 3.7221 & 2.0000 & 4.0152 & 0.3614 \\ \hline
2.9000   & 0.9118   & 2.5586  & 3.8550 & 2.0000 & 4.0852 & 0.3563 \\ \hline
3.0000   & 0.8847   & 2.6511  & 3.9880 & 2.0000 & 4.1528 & 0.3507 \\ \hline
4.0000   & 0.7089   & 3.5448  & 5.3173 & 2.0000 & 4.7284 & 0.3017 \\ \hline
5.0000   & 0.6111   & 4.4311  & 6.6467 & 2.0000 & 5.1744 & 0.2577 \\ \hline
\end{tabular}
\label{table2}
\end{center}
\end{table}

\section{Conclusions}
Light matters not only in technology and photonics. Indeed,
in General Relatiovity and Cosmology it became
an important tool to test the theories and open
a window to the Universe.
The earth- or space-based telescope
not only receive the light to give us
a picture of the Universe, but the analysis of its
red-shift revealed, for instance, the accelerated expansion.  
Gravitational red-shift
and 
the cosmic microwave background radiation
are other examples of the importance of 
electromagnetic phenomena in gravity.
In this paper we have picked up  the classical
connection between light and gravity, namely
the deflection of light in gravitational field
and its lensing. For the ``standard'' metrics, like
Schwarzschild and Schwarzschild-deSitter we have 
generalized some formulae and improved upon the
numerical accuracy of the final results. Our approach
reveals that corrections of the cosmological
constant are small given the present value of this
constant. 

Gravity theory, as many other current physical theory, does not seem
to be complete. Both, on macroscopic level, 
where one of the pressing problems is Dark Matter \cite{DM}
whose matter content is not known,
as well as on the
Planck scale where a Quantum Gravity theory awaits
its global acceptance or discovery.
We have studied the light bending
in a variety of models which correspond 
to one of the above problems.
We have chosen a general relativistic DM metric,
a worm-hole and a dirty Black Hole, both inspired
by noncommutative geometry to derive the formulae of the
deflection angle for light. To this we added a 
Holographic screen metric which is motivated by
considerations at Planck scale.
We paid attention to a global phenomenon emerging
in connection with light motion around naked singularities.
In all the cases we have examined we found that
there is a narrow range of possible parameters
which allow the light to be bound to the source of a naked
singularity.

\end{document}